\documentclass[english,twocolumn,aps,pra]{revtex4-1}
\usepackage[T1]{fontenc}
\usepackage[latin9]{inputenc}
\usepackage{geometry}
\usepackage[active]{srcltx}
\usepackage{amstext}
\usepackage{amssymb}
\usepackage{graphicx}
\usepackage{esint}



\usepackage{babel}
\begin{document}

\title{Enhanced sensing of optomechanically induced nonlinearity by linewidth
suppression and optical bistability in cavity-waveguide systems}

\author{Chun-Wang Liu}

\affiliation{Fujian Key Laboratory of Quantum Information and Quantum Optics and
Department of Physics, Fuzhou University, Fuzhou 350116, People\textquoteright s
Republic of China}

\author{Ye Liu}

\affiliation{Fujian Key Laboratory of Quantum Information and Quantum Optics and
Department of Physics, Fuzhou University, Fuzhou 350116, People\textquoteright s
Republic of China}

\author{Lei Du}

\affiliation{Center for Quantum Sciences and School of Physics, Northeast Normal
University, Changchun 130024, China}

\author{Wan-Jun Su}

\affiliation{Fujian Key Laboratory of Quantum Information and Quantum Optics and
Department of Physics, Fuzhou University, Fuzhou 350116, People\textquoteright s
Republic of China}

\author{Huaizhi Wu}

\affiliation{Fujian Key Laboratory of Quantum Information and Quantum Optics and
Department of Physics, Fuzhou University, Fuzhou 350116, People\textquoteright s
Republic of China}

\author{Yong Li}

\affiliation{Center for Theoretical Physics and School of Science, Hainan University,
Haikou 570228, China}

\affiliation{Synergetic Innovation Center for Quantum Effects and Applications,
Hunan Normal University, Changsha 410081, China}
\begin{abstract}
We study enhanced sensing of optomechanically induced nonlinearity
(OMIN) in a cavity-waveguide coupled system. The Hamiltonian of the
system is anti-PT symmetric with the two involved cavities being dissipatively
coupled via the waveguide. When a weak waveguide-mediated coherent
coupling is introduced, the anti-PT symmetry may break down. However,
we find a strong bistable response of the cavity intensity to the
OMIN near the cavity resonance, benefiting from linewidth suppression
caused by the vacuum induced coherence. The joint effect of optical
bistability and the linewidth suppression is inaccessible by the anti-PT
symmetric system involving only dissipative coupling. Due to that,
the sensitivity is greatly enhanced by two orders of magnitude compared
to that for the anti-PT symmetric model. Moreover, the sensitivity
shows resistances to a reasonably large cavity decay and robustness
to fluctuations in the cavity-waveguide detuning. Based on the integrated
optomechanical cavity-waveguide systems, the scheme can be used for
sensing different physical quantities related to the single-photon
coupling strength, and has potential applications in high-precision
measurements with physical systems involving Kerr-type nonlinearity.
\end{abstract}

\keywords{optomechanics, optical waveguide, anti-PT symmetry, nonlinearity
sensing}
\maketitle

\section{Introduction}

Sensing has played an important role in cutting-edge technologies,
associated with the-state-of-the-art metrology platforms for measurements
of various physical quantities \citep{Degen2017}, ranging from magnetic
and electric fields \citep{Brownnutt2015,Jensen2016}, to time and
frequency \citep{Schmitt2017}, to minute force \citep{Moser2013},
acceleration \citep{Krause2012}, and mass \citep{GilSantos2010,Wiersig2014}.
In contrast to the conventional Hermitian systems, where the sensing
of weak perturbations hinges on the linear response of resonant spectrum
shifts or splitting to a perturbation parameter, recent advances have
found that a new type of sensing schemes can be built on the non-Hermitian
Hamiltonians with Parity-Time (PT) \citep{Chang_NatPhonon2014a,Xiao2019,Dora2021,Katsantonis2022}
or anti-PT symmetry \citep{Ge2013,Wang2022,Luo2022,Park2021,Zhang2020,Nair2021}.
The spectra for such a kind of non-Hermitian Hamiltonians typically
exhibit exceptional point (EP) singularities, at which the eigenvalues
and the corresponding eigenvectors coalesce \citep{Wiersig2014,Chen2017,Hodaei2017,El-Ganainy2018,Miri2019}.
As a result, an external perturbation on the system parameter across
the EPs leads to phase transitions between the PT (anti-PT) phase
with a purely real (imaginary) spectra and the broken PT (anti-PT)
phase with complex eigenvalues. Moreover, the response to linear
perturbations shows a power-law divergence behavior in the first-order
derivative with respect to the perturbation parameter \citep{Wang2022},
which makes the PT and anti-PT symmetric systems the good candidates
for sensitivity enhancement \citep{Li2020,Yang2020,Zhang2020,Peng2016}.
The EP physics has been experimentally demonstrated in different settings
for realizing nonreciprocal light transport \citep{Yin2013,White2011},
single-mode lasers \citep{Feng2014,HosseinHodaeiMohammad-AliMiri2014},
time-asymmetric topological operations \citep{Yoon2018,Zhang2019},
enhanced sensing \citep{Chen2018,Dong2019,Hodaei2017,Xiao2019}, as
well as energy-difference conserving dynamics \citep{Choi2018}.

While early EP studies have focused on sensing linear perturbations
with non-Hermitian systems \citep{Zhang2020}, recent attentions have
been paid to the detection of anharmonic perturbations, which are
essential for quantum state engineering and creation of non-Gaussian
bosonic fields. As an example, Nair \textit{et al}. have recently
proposed an anti-PT symmetry sensing enhancement scheme in the context
of a weakly anharmonic yttrium iron garnet sphere interacting with
a cavity via a fiber waveguide in the vacuum \citep{Nair2021}. In
contrast to the gain-loss balanced system in the PT symmetric phase
\citep{Dora2021,Katsantonis2022,Xiao2019}, which has purely real
eigenvalues, the system in the anti-PT symmetric phase may have a
unique singular point corresponding to the emergence of only one real
eigenvalue, which is referred to as the linewidth suppression caused
by the vacuum induced coherence (VIC) \citep{Paspalakis1998,Keitel1999,Agarwal2000,Scully2010,Heeg2013,Jha2015}.
It was shown that the nonlinear response of the cavity intensity around
the singularity can efficiently quantify the strength of the anharmonicity
explicitly included in the Kittel mode of spins \citep{Hisatomi2016}.
Besides the cavity-magnonics \citep{Yang2020}, many other physical
settings, which have been tailored to exhibit anti-PT symmetry \citep{Konotop2018,Mukhopadhyay2022,Bergman2021,Antonosyan2015}
and to sense linear perturbations \citep{Zhang2020,Luo2022,Li2021},
may have potentials to detect fundamental nonlinear oscillators, manifesting
themselves in many modern fields of physics, e.g., quantum electrical
circuits \citep{Blais2021}, cold atoms \citep{Gothe2019}, levitated
nanoparticles \citep{Fonseca2016,Zheng2020}, and optomechanical systems
\citep{Katz2007,Aspelmeyer2013}

\begin{figure}
\includegraphics[width=1\columnwidth]{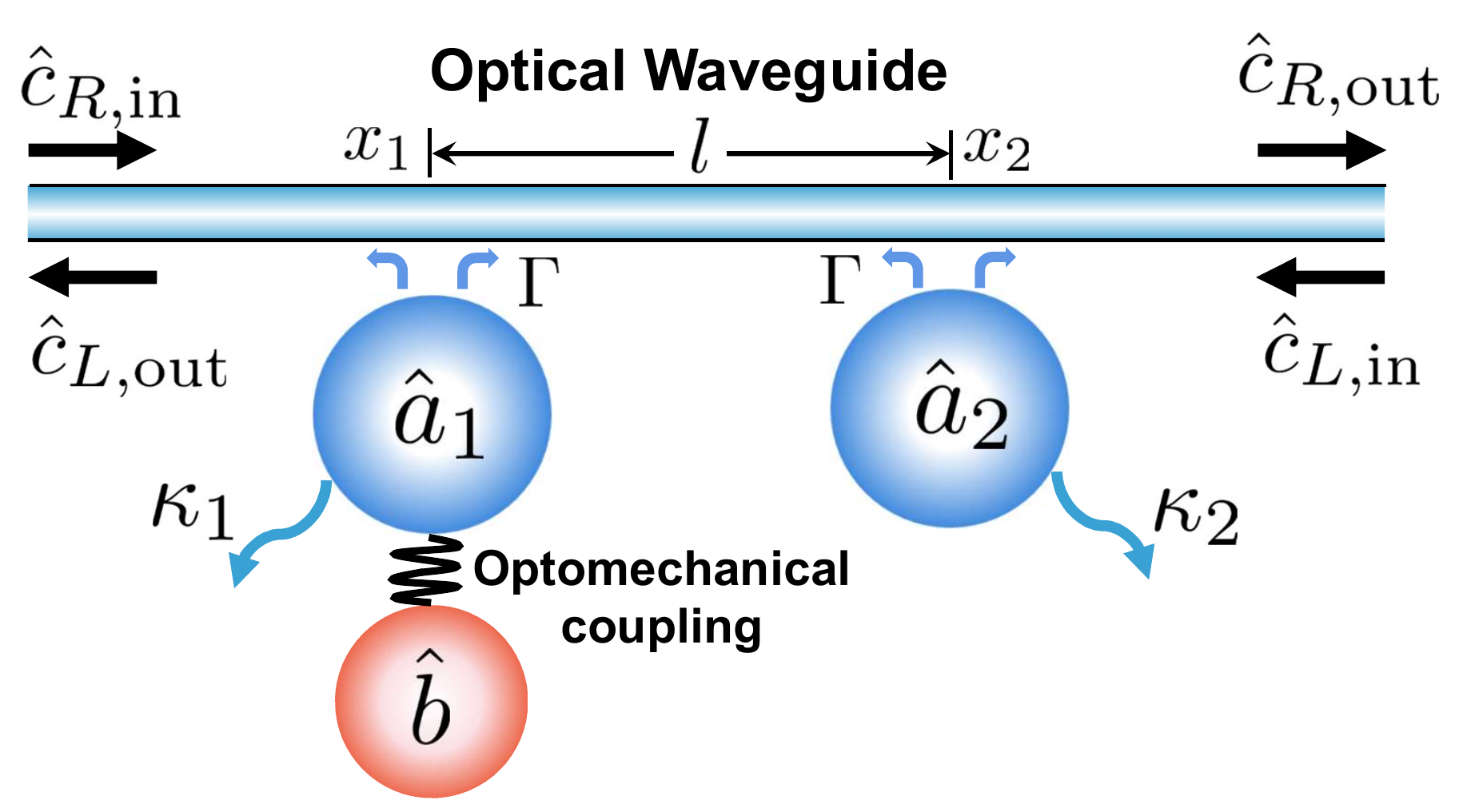}\caption{\label{fig:1}(Color online) Schematic of the optomechanical cavity-waveguide
system. A cavity optomechanical system (consisting of an optical mode
$\hat{a}_{1}$ and a mechanical mode $\hat{b}$) is coupled to the
auxiliary cavity mode $\hat{a}_{2}$ through an optical waveguide
with a linear dispersion relation. $\hat{c}_{L,\text{in}}(x_{j},t)$
and $\hat{c}_{R,\text{in}}(x_{j},t)$ denote the left- and right-moving
fields which interact with the cavity modes at the position $x_{j}$
with the strength $\Gamma$. The cavity modes $\hat{a}_{j}$ decay
into the surrounding environments with the rates $\kappa_{j}$. }
\end{figure}

In this paper, we study enhanced sensing of optomechanical interactions
with a coupled cavity-waveguide system. The system consists of an
optomechanical cavity and an auxiliary cavity, which are coupled via
a one-dimensional waveguide in the vacuum. The Hamiltonian of the
system can be anti-PT symmetric if the two cavities are dissipatively
coupled, which is valid only when the phase accumulation of light
propagation from one cavity to the other is precisely a multiple of
$2\pi$. When a phase deviation is introduced, a coherent coupling
between the cavity modes is induced and the anti-PT symmetry can not
hold. Then, there does not exist the eigenvalue with vanishing imaginary
part, and the eigenmodes are subject to a finite loss to the waveguide,
which may be detrimental to sensing of anharmonicity in the cavity
modes \citep{Nair2021}. On the other hand, it is well known that
optical bistability can appear in the presence of a Kerr nonlinearity
with appropriate cavity drivings. However, the standard anti-PT symmetric
system shows bistability only when the frequency detuning between
the cavity modes is large enough. This condition is related to the
broken anti-PT phase, where the VIC induced linewidth suppression
can not be found \citep{Nair2021b}. Here, by considering the optomechanically
induced Kerr-type nonlinearity and introducing a phase deviation to
the field propagation, we find that optical bistability can occur
at the cavity resonance, where the system Hamiltonian does not have
the purely real eigenvalue, but the VIC induced linewidth suppression
still holds, corresponding to a long-lived eigenmode. By combining
the advantages of the resonant linewidth suppression and the optical
bistability, a strong bistable response of the cavity intensity to
the OMIN can be found. By comparing the responses to two optomechanical
strengths with a factor of 3 difference, we identify two working regions
for sensing of the OMIN. In the first region, the system may display
monostability (or optical bistability) for both coupling strengths,
where the sensitivity can be double compared to that of the anti-PT
symmetric model. In the second region, the system displays monostable
behavior for one strength and optical bistability for the other, then
a remarkably high sensitivity of a few hundreds can be achieved. Moreover,
the sensing scheme is resistant to a weak cavity decay, and effectively
functions with the sensitivity remaining a few tens. As an example,
we envision a cascaded optomechanical setup where an optically levitated
nanodiamond couples to two waveguide-coupled microspheres \citep{Juan2016}.
The model can be potentially generalized to other integrated optomechanical
cavity-waveguide systems, such as hybrid cavity-magnonic systems \citep{Harder2018,ZareRameshti2018},
optical crystal circuits \citep{Fang2016}, and microwave optomechanical
circuits \citep{Bernier2018}. 

The remainder of the paper is organized as follows. In Sec. II, we
first derive the coherent and dissipative couplings between two optical
cavities coupled via a vacuum waveguide. In Sec. III, we recall the
anti-PT symmetric system constructed by the coupled cavity-waveguide
setup, and discuss the VIC induced linewidth suppression significantly
for sensing of the OMIN. In Sec. IV, we examine the coherence induced
bistability around the cavity resonance, based on which, the sensitivity
enhancement can be realized, as studied in Sec. V. We discuss the
experimental feasibility of the scheme in Sec. VI and summarized our
results in Sec. VII.

\section{Theoretical Model}

As schematically shown in Fig. \ref{fig:1}, we start by considering
the general model where two cavity modes $\hat{a}_{1}$ and $\hat{a}_{2}$
of frequencies $\omega_{1}$ and $\omega_{2}$, respectively, are
coupled via a waveguide, and have the decays rates $\kappa_{1}$ and
$\kappa_{2}$. The cavity mode $\hat{a}_{1}$ is coupled to a mechanical
mode of the frequency $\omega_{m}$ via the standard optomechanical
interaction. For simplicity, we assume that the waveguide has a linear
dispersion relation $\omega(k)=v_{G}|k|$ for the left- and right-going
waveguide modes of wavevector $k$. Working at the rotating frame
with respect to the central frequency $\omega_{0}\equiv(\omega_{1}+\omega_{2})/2$
of the two coupled cavities, the Hamiltonian of the system is then
given by ($\hbar=1$) \citep{Metelmann2015,Chang2011,Wang2019,Du2021}
\begin{equation}
H=H_{C}+H_{W}+H_{CW}
\end{equation}
with

\begin{eqnarray}
H_{C} & = & \frac{\delta}{2}\left(\hat{a}_{1}^{\dagger}\hat{a}_{1}-\hat{a}_{2}^{\dagger}\hat{a}_{2}\right)+\frac{\omega_{m}}{2}(\hat{p}^{2}+\hat{q}^{2})+g\hat{a}_{1}^{\dagger}\hat{a}_{1}\hat{q}\nonumber \\
 &  & +i\Omega\hat{a}_{1}^{\dagger}e^{-i(\omega_{d}-\omega_{0})t}+\text{H.c.}\text{,}
\end{eqnarray}
\begin{eqnarray}
H_{W} & = & \int dx[\hat{c}_{R}^{\dagger}(x)(-iv_{G}\frac{\partial}{\partial x}-\omega_{0})\hat{c}_{R}(x)\nonumber \\
 &  & +\hat{c}_{L}^{\dagger}(x)(iv_{G}\frac{\partial}{\partial x}-\omega_{0})\hat{c}_{L}(x)],
\end{eqnarray}
\begin{equation}
H_{CW}=-\sqrt{\frac{\Gamma v_{G}}{2}}\sum_{j=1,2}\hat{a}_{j}^{\dagger}[\hat{c}_{R}(x_{j})+\hat{c}_{L}(x_{j})]+\text{H.c.},
\end{equation}
where $H_{C}$ describes the interaction of the cavity mode with the
mechanical motion via the radiation pressure force with $\hat{q}$
($\hat{p}$) being the mechanical displacement (momentum) operator,
and $g$ being the single-photon optomechanical coupling strength.
The cavity mode $a_{1}$ is driven by an external laser field of the
amplitude $\Omega=\sqrt{P_{\text{in}}\kappa_{1}/\hbar\omega_{d}}$
with $P_{\text{in}}$ being the input power and $\omega_{d}$ the
laser frequency, which is set to $\omega_{d}=\omega_{0}$ in what
follows. $\hat{c}_{L,R}(x_{j})$ denote the left- and right-going
modes in the waveguide, which couple to cavity $j$ at position $x_{j}$
with the rate $\Gamma$ \citep{Metelmann2015}.

The optical cavities can be dissipatively coupled by reservoir engineering
the waveguide field, and can be further designed as an anti-PT symmetric
system \citep{Nair2021}. To see this, we first leave the optomechanical
interaction out and include it later as the probe. Then, the Heisenberg
equations of motion for the waveguide modes are given by \citep{Chang2011}
\begin{eqnarray}
\frac{\partial}{\partial t}\hat{c}_{R}(x,t) & = & (-v_{G}\frac{\partial}{\partial x}+i\omega_{0})\hat{c}_{R}(x,t)\nonumber \\
 &  & +i\sum_{j=1,2}\sqrt{\frac{\Gamma v_{G}}{2}}\delta(x-x_{j})\hat{a}_{j},\\
\frac{\partial}{\partial t}\hat{c}_{L}(x,t) & = & (v_{G}\frac{\partial}{\partial x}+i\omega_{0})\hat{c}_{L}(x,t)\nonumber \\
 &  & +i\sum_{j=1,2}\sqrt{\frac{\Gamma v_{G}}{2}}\delta(x-x_{j})\hat{a}_{j}.
\end{eqnarray}
By integrating the field equations across the discontinuity at $x_{j}$,
one obtains the input-output relations between the cavity modes and
the left- (right-) going fields \citep{Chang2011}
\begin{eqnarray}
\hat{c}_{R}(x_{j}^{+},t)-\hat{c}_{R}(x_{j}^{-},t) & = & i\sqrt{\frac{\Gamma}{2v_{G}}}\hat{a}_{j}(t),\label{eq:R+-}
\end{eqnarray}
\begin{eqnarray}
\hat{c}_{L}(x_{j}^{-},t)-\hat{c}_{L}(x_{j}^{+},t) & = & i\sqrt{\frac{\Gamma}{2v_{G}}}\hat{a}_{j}(t).\label{eq:L+-}
\end{eqnarray}
For clarity, we can further define $\hat{c}_{L,\text{in}}(x_{j},t)\equiv\hat{c}_{L}(x_{j}^{+},t)$
and $\hat{c}_{R,\text{in}}(x_{j},t)\equiv\hat{c}_{R}(x_{j}^{-},t)$
as the input fields to the cavities from the waveguide, and $\hat{c}_{L,\text{out}}(x_{j},t)\equiv\hat{c}_{L}(x_{j}^{-},t)$
and $\hat{c}_{R,\text{out}}(x_{j},t)\equiv\hat{c}_{R}(x_{j}^{+},t)$
as the output fields. This leads to the familiar form of the standard
input-output relations:
\begin{eqnarray}
\hat{c}_{R,\text{out}}(x_{j},t) & = & \hat{c}_{R,\text{in}}(x_{j},t)+i\sqrt{\frac{\Gamma}{2v_{G}}}\hat{a}_{j},\nonumber \\
\hat{c}_{L,\text{out}}(x_{j},t) & = & \hat{c}_{L,\text{in}}(x_{j},t)+i\sqrt{\frac{\Gamma}{2v_{G}}}\hat{a}_{j}.\label{eq:input-output relation}
\end{eqnarray}
In addition, since the waveguide field propagates freely between the
cavities and the waveguide dispersion is linear over all frequencies,
we have \citep{Xiao2010,Metelmann2015,Du2021oe}

\begin{eqnarray}
\hat{c}_{R,\text{in}}(x_{2},t) & = & e^{i\Phi}\hat{c}_{R,\text{out}}(x_{1},t),\nonumber \\
\hat{c}_{L\text{,in}}(x_{1},t) & = & e^{i\Phi}\hat{c}_{L,\text{out}}(x_{2},t)\text{,}\label{eq:field_prop}
\end{eqnarray}
where $\Phi=kl$ with $l\equiv|x_{2}-x_{1}|$. Consider the case where
for the frequencies of interest around $\omega_{0}$, the phase delay
$\delta(l/v_{G})\ll1$ due to field propagation is ignorable, we can
then replace $k$ by $k_{0}\equiv\omega_{0}/v_{G}$ \citep{Metelmann2015}.
Moreover, we are concerned with the Markovian approximation $\kappa(l/v_{G})\ll1$,
and thus the retarded effect on the cavity dynamics due to the field
propagation can also be neglected. Based on these assumptions, one
can achieve the $\Phi$-dependent coupling between the two cavities.
To see this, we further write down the Heisenberg equations of motion
for the cavity operators without the optomechanical coupling,

\begin{eqnarray}
\dot{\hat{a}}_{1} & = & -\left(-i\frac{\delta}{2}+\frac{\kappa_{1}}{2}\right)\hat{a}_{1}+\Omega+\sqrt{\kappa_{1}}\hat{a}_{1,\text{in}}\nonumber \\
 &  & +i\sqrt{\frac{\Gamma v_{G}}{2}}[\hat{c}_{R,\text{out}}(x_{1},t)+\hat{c}_{L,\text{in}}(x_{1},t)],\label{eq:EOM_a1}\\
\dot{\hat{a}}_{2} & = & -\left(i\frac{\delta}{2}+\frac{\kappa_{2}}{2}\right)\hat{a}_{2}+\sqrt{\kappa_{2}}\hat{a}_{2,\text{in}}\nonumber \\
 &  & +i\sqrt{\frac{\Gamma v_{G}}{2}}[\hat{c}_{R,\text{out}}(x_{2},t)+\hat{c}_{L,\text{in}}(x_{2},t)],\label{eq:EOM_a2}
\end{eqnarray}
and substitute Eqs. (\ref{eq:input-output relation}) and (\ref{eq:field_prop})
into Eqs. (\ref{eq:EOM_a1}) and (\ref{eq:EOM_a2}). It then follows
that 

\begin{eqnarray}
\dot{\hat{a}}_{1} & = & -\left(-i\frac{\delta}{2}+\frac{\kappa_{1}+\Gamma}{2}\right)\hat{a}_{1}-ie^{i\Phi}\frac{\Gamma}{2}\hat{a}_{2}+\Omega+\sqrt{\kappa}\hat{a}_{1,\text{in}}\nonumber \\
 &  & +i\sqrt{\frac{\Gamma v_{G}}{2}}\left[\hat{c}_{R,\text{in}}(x_{1},t)+e^{i\Phi}\hat{c}_{L,\text{in}}(x_{2},t)\right],\nonumber \\
\\
\dot{\hat{a}}_{2} & = & -\left(i\frac{\delta}{2}+\frac{\kappa_{2}+\Gamma}{2}\right)\hat{a}_{2}-ie^{i\Phi}\frac{\Gamma}{2}\hat{a}_{1}+\sqrt{\kappa}\hat{a}_{2,\text{in}}\nonumber \\
 &  & +i\sqrt{\frac{\Gamma v_{G}}{2}}\left[e^{i\Phi}\hat{c}_{R,\text{in}}(x_{1},t)+\hat{c}_{L,\text{in}}(x_{2},t)\right],\nonumber \\
\end{eqnarray}
where $\hat{a}_{j,\text{in}}$ ($j=1,2$) are zero-mean quantum noises
and the cavity decay rates $\kappa_{1}=\kappa_{2}=\kappa$ are assumed
to be identical. The above equations can be rewritten in the matrix
form
\begin{eqnarray}
\left(\begin{array}{c}
\dot{\hat{a}}_{1}\\
\dot{\hat{a}}_{2}
\end{array}\right) & = & -i\mathcal{H}\left(\begin{array}{c}
\hat{a}_{1}\\
\hat{a}_{2}
\end{array}\right)+\left(\begin{array}{c}
\Omega+\hat{a}_{1,N}\\
\hat{a}_{2,N}
\end{array}\right),\label{eq:Matrix langevin}
\end{eqnarray}
with
\begin{equation}
\mathcal{H}=\left(\begin{array}{cc}
-\frac{\delta}{2}-i\frac{\kappa+\text{\ensuremath{\Gamma}}}{2} & -i\frac{\Gamma}{2}\text{cos}\Phi+\frac{\Gamma}{2}\text{sin}\Phi\\
-i\frac{\Gamma}{2}\text{cos}\Phi+\frac{\Gamma}{2}\text{sin}\Phi & \frac{\delta}{2}-i\frac{\kappa+\Gamma}{2}
\end{array}\right)
\end{equation}
and the total input noise operators being defined as

\begin{eqnarray}
\hat{a}_{1,N} & = & \sqrt{\kappa}\hat{a}_{1,\text{in}}+i\sqrt{\frac{\Gamma v_{G}}{2}}\left[\hat{c}_{R,\text{in}}(x_{1},t)+e^{i\Phi}\hat{c}_{L,\text{in}}(x_{2},t)\right],\nonumber \\
\hat{a}_{2,N} & = & \sqrt{\kappa}\hat{a}_{2,\text{in}}+i\sqrt{\frac{\Gamma v_{G}}{2}}\left[e^{i\Phi}\hat{c}_{R,\text{in}}(x_{1},t)+\hat{c}_{L,\text{in}}(x_{2},t)\right],\nonumber \\
\end{eqnarray}
which have the zero mean for $\hat{c}_{L,\text{in}}(x_{1},t)$ and
$\hat{c}_{R,\text{in}}(x_{1},t)$ being the Gaussian white noises
as well \citep{Metelmann2015}. Note that, the waveguide acts as an
engineered reservoir, which induces coherent and dissipative couplings
between the two optical modes with amplitudes $(\Gamma/2)\text{sin}\Phi$
and $(\Gamma/2)\text{cos}\Phi$, respectively. The model, under the
anti-PT symmetric case (i.e. $e^{i\Phi}=\pm1$), has been applied
for sensing weak anharmonicities in a yttrium iron garnet sphere interacting
with a cavity-fiber system \citep{Nair2021}. Instead of placing a
Kerr interaction term in the Hamiltonian, here in our case, the optomechanical
interaction with the single-photon coupling strength $g$ is included.
A Kerr-type nonlinearity proportional to $g^{2}$ in the optomechanical
cavity will be generated when the system is in the steady state, see
details in the following section. We then focus on the enhanced sensing
of weak optomechanical interactions by using the coupled cavity-waveguide
system with a finite coherent coupling (i.e. $e^{i\Phi}\neq\pm1$),
which can be efficient for measurement of the physical quantities
related to the optomechanical single-photon coupling strength $g$.

\section{The VIC induced linewidth suppression }

\begin{figure}
\includegraphics[width=1\columnwidth]{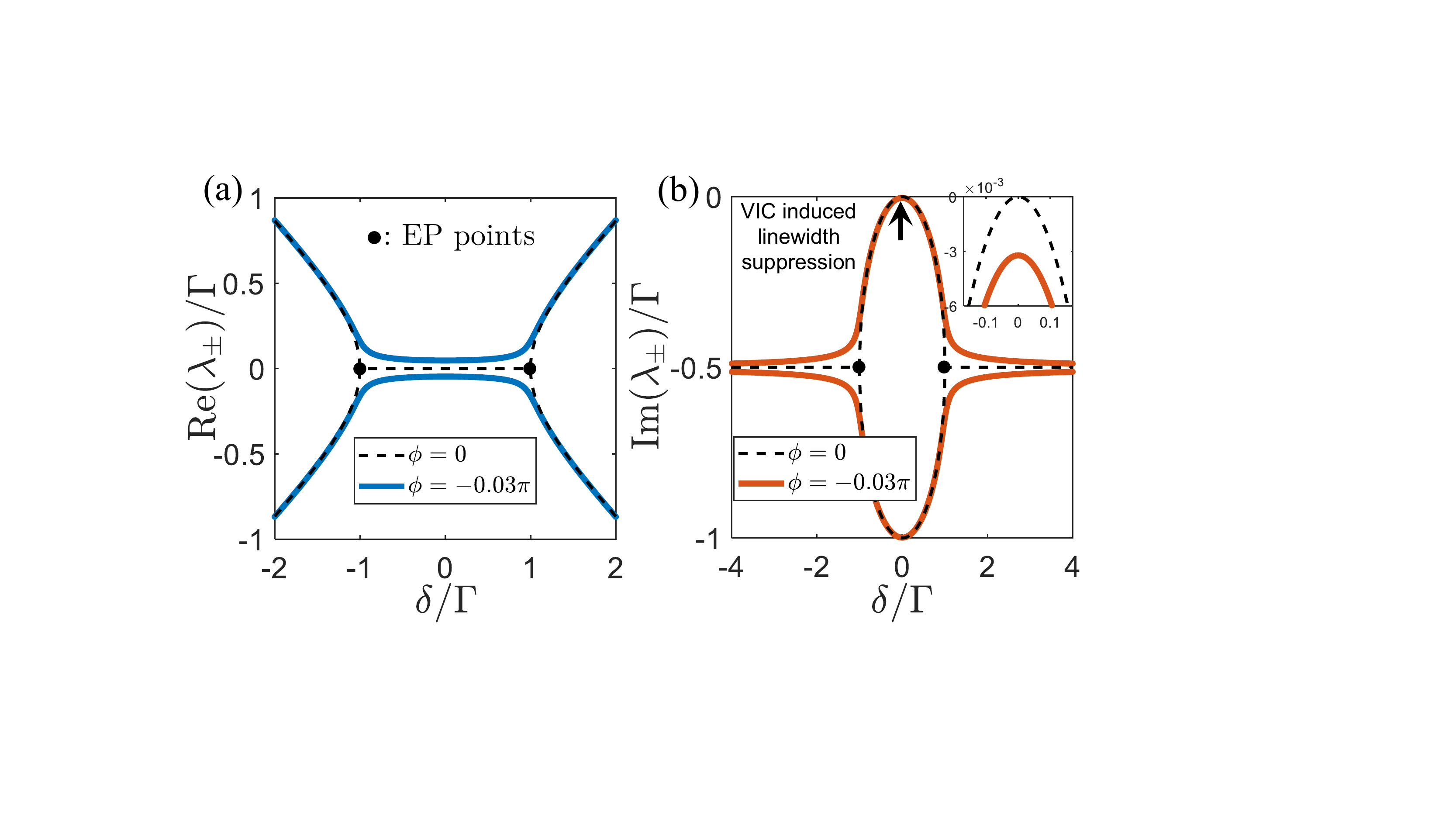}\caption{\label{fig:2}(Color online) (a) Real parts (eigenfrequencies) and
(b) imaginary parts (linewidths) of the eigenvalues of $\mathcal{H}$
versus the detuning $\delta/\Gamma$ for $\phi=0$ (dashed) and $\phi=-0.03\pi$
(solid), respectively. The cavity decay rates are set to zero such
that the VIC induced zero linewidth ($\text{Im\ensuremath{\lambda_{+}/\Gamma=0}}$)
arises at the resonance ($\delta=0$) for the anti-PT symmetric regime
$\phi=0$. The EPs disappear for $\phi\protect\neq0$, but the VIC
induced linewidth suppression (with $\text{Im\ensuremath{\lambda_{+}/\Gamma\rightarrow10^{-3}}}$)
remains effective for $\delta\rightarrow0$ and $\phi\ll\pi$, as
can be seen from the inset. Here, we consider a finite cavity linewidth
with $\kappa/\Gamma=0.002$.}
\end{figure}
For the propagating phase being a multiple of $2\pi$, i.e. $\Phi=2n\pi$
($n\in\mathit{Z}$), the coherent coupling with the amplitude $(\Gamma/2)\text{sin}\Phi$
between the cavity modes vanishes \citep{Metelmann2015}, leading
to a purely dissipative hopping interaction of the strength $\Gamma/2$.
The matrix $\mathcal{H}$ reduces to the standard anti-PT symmetric
form
\begin{eqnarray}
\mathcal{H}_{APT} & = & \left(\begin{array}{cc}
-\frac{\delta}{2}-i\frac{\kappa+\text{\ensuremath{\Gamma}}}{2} & -i\frac{\Gamma}{2}\\
-i\frac{\Gamma}{2} & \frac{\delta}{2}-i\frac{\kappa+\Gamma}{2}
\end{array}\right),\label{eq:H_antiPT}
\end{eqnarray}
 where the two coupled cavity modes have: (i) opposite frequency detunings
and the same gain or loss; and (ii) dissipative coupling (anti-Hermitian
coupling) between them. Thus, the Hamiltonian obeys $(\widehat{PT})\mathcal{H}_{APT}(\widehat{PT})=-\mathcal{H}_{APT}$
and is anti-PT symmetric \citep{Yang2020,Nair2021}. One can readily
obtain the eigenvalues of $\mathcal{H}_{APT}$,

\begin{equation}
\lambda_{\pm}=-i\frac{\kappa+\Gamma}{2}\pm\frac{1}{2}\sqrt{\delta^{2}-\Gamma^{2}},\label{eq:eigenvalues}
\end{equation}
which are plotted against $\delta/\Gamma$, as shown in Fig. \ref{fig:2}.
In the anti-PT symmetric phase $\left|\delta/\Gamma\right|<1$, $\mathcal{H}_{APT}$
has purely imaginary eigenvalues $\lambda_{\pm}=-i\frac{\kappa+\Gamma}{2}\pm i\frac{1}{2}\lambda_{0}$
with $\lambda_{0}=\sqrt{|\delta^{2}-\Gamma^{2}|}$. The symmetry breaking
occurs at the exceptional points (EPs) $\left|\delta/\Gamma\right|=1$,
where the eigenstates coalesce, $\lambda_{\pm}=-i\frac{\kappa+\Gamma}{2}$.
In the broken anti-PT phase $\left|\delta/\Gamma\right|>1$, the two
eigenvalues $\lambda_{\pm}=-i\frac{\kappa+\Gamma}{2}\pm\frac{1}{2}\lambda_{0}$
are complex. For the cavity decay rate $\kappa=0$, there exists the
singular point ($\delta=0$) corresponding to the VIC induced zero
linewidth (namely $\text{Im}\lambda_{+}=0$) in the anti-PT symmetric
phase. The feature is essential for sensing of weak anharmonicities
\citep{Nair2021}.

For $\Phi=2n\pi+\phi$ with a finite phase deviation $\phi\neq0$,
the matrix $\mathcal{H}$ becomes $\phi$ dependent. A weak coherent
coupling with the amplitude $(\Gamma/2)\text{sin}\Phi\sim\Gamma\phi/2$
for $\phi\ll\pi$ is introduced to the coupled cavity modes, and the
anti-PT symmetry breaks down. The eigenvalues of the matrix $\mathcal{H}$
now read

\begin{equation}
\lambda_{\pm}=-i(\frac{\kappa+\Gamma}{2}\mp\frac{\Gamma}{2}\text{\ensuremath{\tilde{\lambda}_{0}}cos\ensuremath{\frac{\theta}{2}}})\mp\frac{\Gamma}{2}\tilde{\lambda}_{0}\text{sin\ensuremath{\frac{\ensuremath{\theta}}{2}}}
\end{equation}
with

\begin{equation}
\tilde{\lambda}_{0}=\ensuremath{\left(1+\frac{\delta^{4}}{\Gamma^{4}}-2\frac{\delta^{2}}{\Gamma^{2}}\text{cos}2\Phi\right){}^{\frac{1}{4}}},
\end{equation}
\begin{equation}
\theta=\text{arctan}\left(\frac{\Gamma^{2}\text{sin\ensuremath{2\Phi}}}{\Gamma^{2}\text{cos}2\Phi-\delta^{2}}\right).
\end{equation}
Compared with the case of the completely dissipative coupling regime,
both the real and imaginary parts of $\lambda_{\pm}$ become nondegenerate
and nonzero regardless of the value of $\delta$, see the solid lines
in Figs. \ref{fig:2}(a)-(b). In other words, there does not exist
the EPs which strictly correspond to the degeneracy points in the
spectra. In this case, we note that $\theta=2\phi$ at $\delta=0$,
and thus $\text{Im}\lambda_{+}\approx-\frac{1}{2}(\kappa+\frac{\Gamma}{2}\phi^{2})$,
as indicated by the arrow in Fig. \ref{fig:2}(b). It implies that
the system is now subject to a weak loss induced by the waveguide-mediated
coherent coupling, namely, there does not exist the real singularity
with a completely vanishing linewidth (i.e. the VIC induced zero linewidth)
even though the cavity decay $\kappa$ is set to zero. When the system
is engineered to be anti-PT symmetric, a dissipative eigenmode may
reduce the sensitivity for sensing of weak anharmonicity in the cavity
modes \citep{Nair2021}. However, the imaginary part of $\lambda_{+}$
remains negligible compared to $\Gamma$ (i.e. $\text{Im}\lambda_{+}/\Gamma\sim10^{-3}$),
as the result of the coherence induced symmetry breaking, see the
inset in Fig. \ref{fig:2}(b). Indeed, we find that a weak coherent
coupling $(\Gamma/2)\text{sin}\phi$ still allows for a long-lived
cavity resonance, and remarkably, it can induce optical bistability
around the cavity resonance at weak cavity driving power, which can
not happen for the anti-PT symmetric Hamiltonian Eq. (\ref{eq:H_antiPT})
with $\delta=0$ \citep{Nair2021b}. By combining the linewidth suppression
with the optical bistability, the sensitivity of detecting the OMIN
can be greatly enhanced, and moreover, can be resistant to a finite
inherent cavity linewidth, see the details in Sec. V.

\section{Coherence induced bistability }

To demonstrate enhanced sensing of the OMIN with the coupled cavity-waveguide
system, we now include the optomechanical coupling term as a perturbative
probe, and consider the classical Langevin equation of motions for
the mean (steady-state) values of the field and mechanical operators
$\alpha_{j}\equiv\langle\hat{a}_{j}\rangle$, $\langle\hat{q}\rangle$
and $\langle\hat{p}\rangle$ under the mean-field approximation $\langle\hat{q}\hat{a}_{1}\rangle\approx\langle\hat{q}\rangle\langle\hat{a}_{1}\rangle$
and $\langle\hat{a}_{1}^{\dagger}\hat{a}_{1}\rangle\approx|\alpha_{1}|^{2}$.
It follows that
\begin{eqnarray}
\dot{\alpha}_{1} & = & i\frac{\delta}{2}\alpha_{1}-\frac{\kappa+\Gamma}{2}\alpha_{1}-e^{i\Phi}\frac{\Gamma}{2}\alpha_{2}\nonumber \\
 &  & -ig\langle\hat{q}\rangle\alpha_{1}+\Omega,\label{eq:alpha1_t}\\
\dot{\alpha}_{2} & = & -i\frac{\delta}{2}\alpha_{2}-\frac{\kappa+\Gamma}{2}\alpha_{2}-e^{i\Phi}\frac{\Gamma}{2}\alpha_{1},\label{eq:alpha2_t}\\
\langle\dot{\hat{q}}\rangle & = & \omega_{m}\langle\hat{p}\rangle,\label{eq:q_t}\\
\langle\dot{\hat{p}}\rangle & = & -\omega_{m}\langle\hat{q}\rangle-g\left|\alpha_{1}\right|^{2}-\gamma_{m}\langle\hat{p}\rangle.\label{eq:p_t}
\end{eqnarray}
In the steady state for Eqs. (\ref{eq:alpha1_t})-(\ref{eq:p_t}),
i.e., $\dot{\alpha}_{1}=\dot{\alpha}_{2}=\dot{\langle\hat{q}\rangle}=\dot{\langle\hat{p}\rangle}=0$,
and by eliminating the mechanical degree of freedom, we find the modified
relations for the field amplitudes
\begin{equation}
(-i\frac{\delta}{2}+\frac{\kappa+\Gamma}{2})\alpha_{1}+e^{i\Phi}\frac{\Gamma}{2}\alpha_{2}-i\frac{g^{2}}{\omega_{m}}\left|\alpha_{1}\right|^{2}\alpha_{1}=\Omega,\label{eq:alpha1_cubic}
\end{equation}
\begin{eqnarray}
\alpha_{2} & = & \frac{-\Gamma e^{i\Phi}}{i\delta+(\kappa+\text{\ensuremath{\Gamma}})}\alpha_{1}.\label{eq:alpha2}
\end{eqnarray}
Moreover, by inserting Eq. (\ref{eq:alpha2}) into Eq. (\ref{eq:alpha1_cubic}),
the cavity intensity $\beta\equiv\left|\alpha_{1}\right|^{2}$ is
found to satisfy a cubic relation

\begin{equation}
\chi^{2}\beta^{3}+\chi A\beta^{2}+B\beta=I,\label{eq:beta_cubic}
\end{equation}
where $\chi=g^{2}/\omega_{m}$ is the Kerr-type nonlinearity induced
by the optomechanical interaction, $I=\Omega^{2}$ is proportional
to the input laser power $P_{\text{in}}$,
\begin{equation}
A=\delta+\frac{\Gamma^{2}\left[-\delta\text{cos}2\Phi+(\kappa+\Gamma)\text{sin}2\Phi\right]}{\delta^{2}+(\kappa+\Gamma)^{2}}
\end{equation}
 and 
\begin{equation}
B=\frac{\delta^{2}+(\kappa+\Gamma)^{2}}{4}+\frac{1}{4}\frac{\Gamma^{4}}{\delta^{2}+(\kappa+\Gamma)^{2}}-\frac{\Gamma^{2}}{2}\text{cos}2\Phi.
\end{equation}
\begin{figure}
\includegraphics[width=1\linewidth]{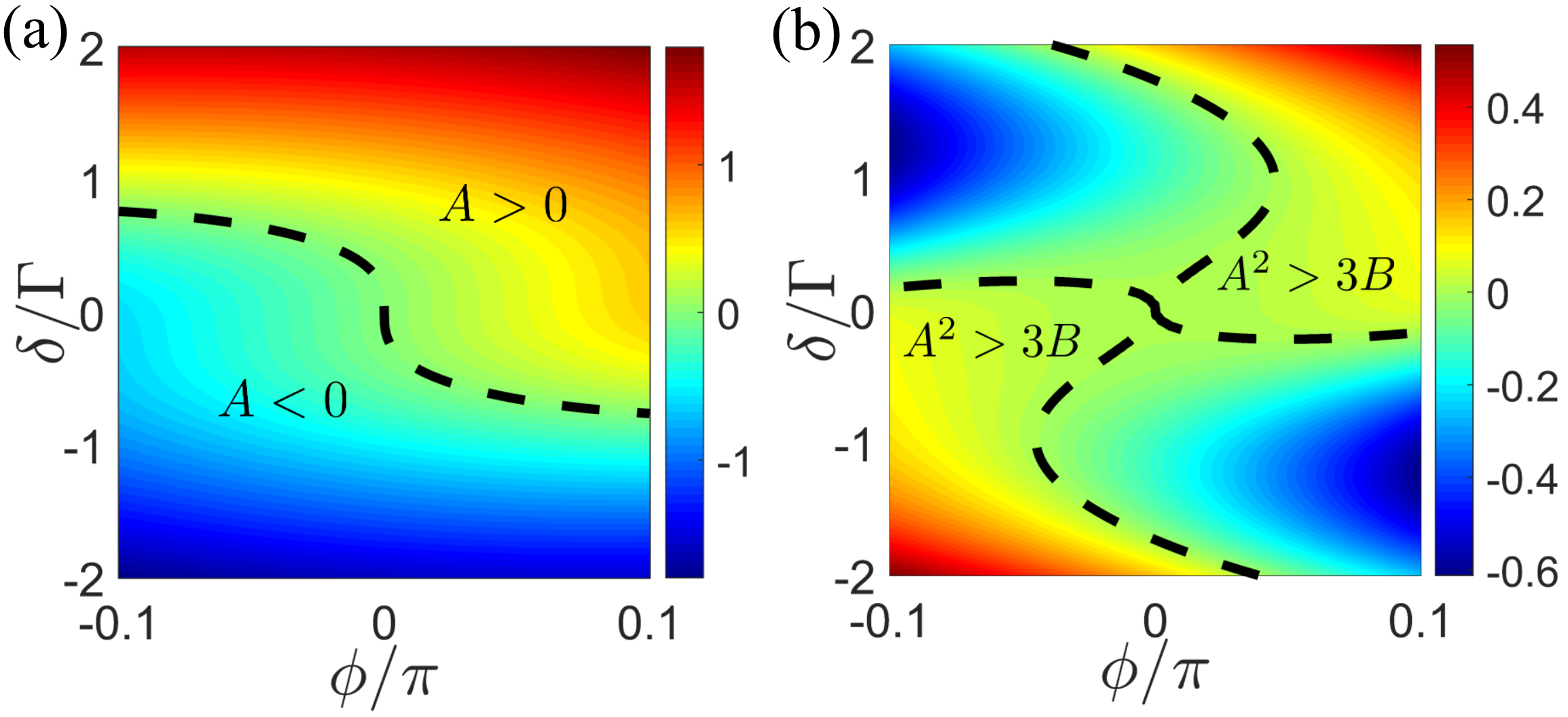}

\caption{\label{StableCondition}(Color online) The necessary conditions for
bistability. (a) Dimensionless $A/\Gamma$ and (b) \textbf{$\left(A^{2}-3B\right)/\Gamma$}
plotted against $\phi$ and $\delta$. The black dashed lines indicate
$A=0$ in (a) and $A^{2}=3B$ in (b).}
\end{figure}

For the completely dissipative coupling regime $\Phi=2n\pi$, Eq.
(\ref{eq:beta_cubic}) entails a bistable response to a strong driving
$I$ only when $|\delta|>\sqrt{3}(\kappa+\Gamma)$, namely, a bistability
behavior can not be found in the anti-PT symmetry phase corresponding
to $|\delta/\Gamma|<1$ . However, we find that a bistable response
to both $\phi$ and $\delta$ can happen by including the weak coherent
coupling $\sim\frac{\Gamma}{2}\text{sin}\Phi$, at low driving power
and small detuning $|\delta/\Gamma|\ll1$. For the driving power dependence
of the cavity intensity $\beta$, the bistability turning points can
be derived from Eq. (\ref{eq:beta_cubic}) by inspecting the solutions
of $\text{d}I/\text{d}\beta=0$, which turn out to be

\begin{eqnarray}
\text{\ensuremath{\beta_{\pm}}} & = & \frac{-A\pm\sqrt{A^{2}-3B}}{3\chi}.\label{eq:bistable condition}
\end{eqnarray}
Thus, the necessary conditions for observing a bistable signature
are $A<0$ and $A^{2}>3B$, which are shown in Fig. \ref{StableCondition}
by plotting them against $\delta$ and $\phi$. The boundaries $A=0$
and $A^{2}=3B$ between the monostable and the bistable regions are
indicated by the dashed lines. It can be seen that the bistable region
embraces the small detuning regime $|\delta/\Gamma|\rightarrow0$
for a finite negative phase $\phi<0$, namely, the VIC induced linewidth
suppression and the static optical bistability can co-exist. By combining
the two effects, the response of the cavity intensity $\beta$ to
optomechanical interactions around $|\delta/\Gamma|\ll1$ can be strong,
which is important for the sensing scheme. However for $\phi>0$,
only a monostable behavior can be found. 

\section{Sensing optomechanical interaction enhanced by bistability}

\begin{figure}
\includegraphics[width=1\linewidth]{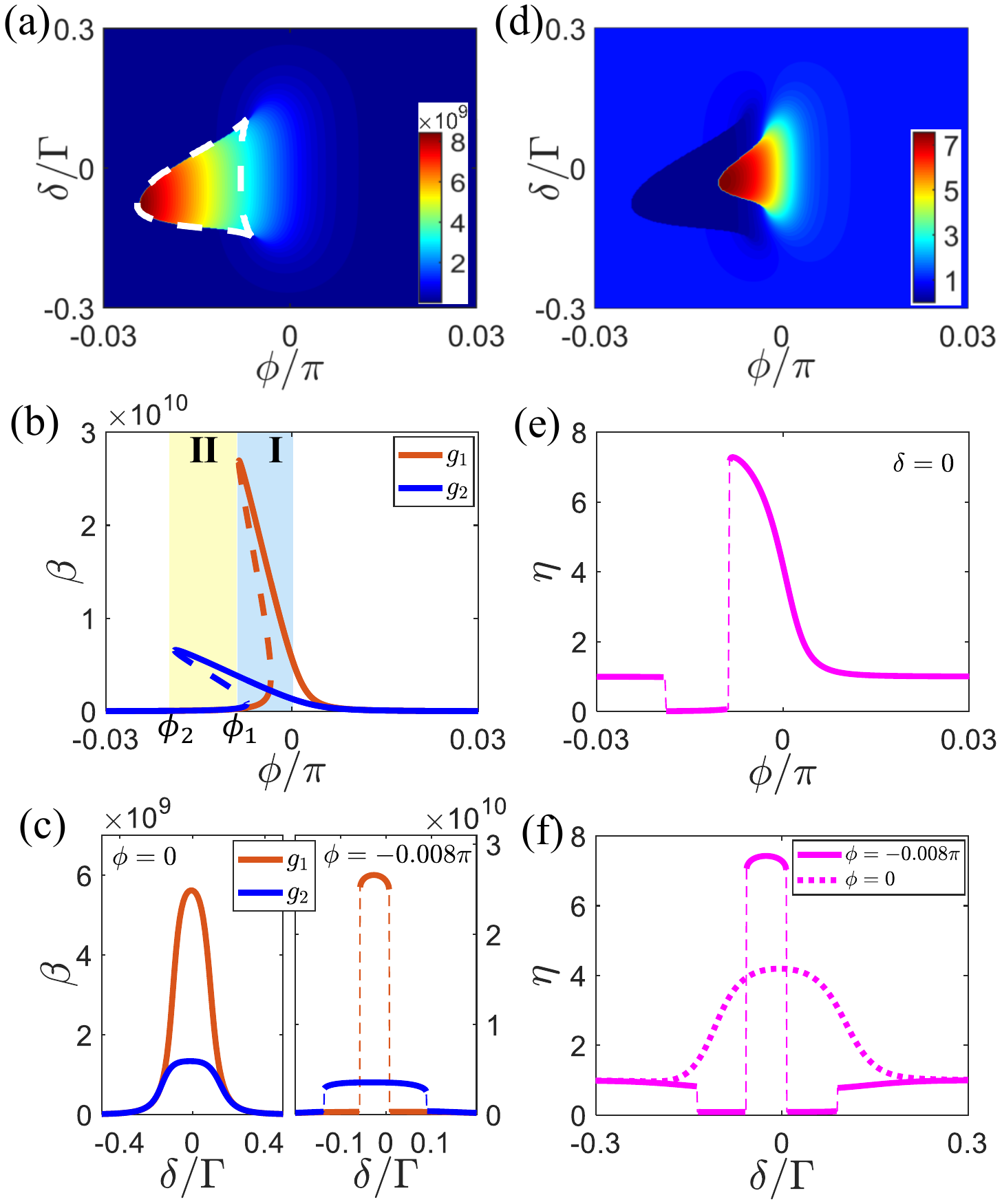}

\caption{\label{fig:4}(Color online) (a) Cavity intensity $\text{\ensuremath{\beta}}$
plotted against $\phi$ and $\delta$ for optomechanical coupling
strength $g/2\pi=3$ Hz. The white dashed line encircles the bistable
region. (b) Cavity intensity $\text{\ensuremath{\beta}}$ on resonance
$\delta=0$ versus $\phi$ for optomechanical coupling strengths $g_{1}/2\pi=1$
Hz and $g_{2}/2\pi=3$ Hz, respectively. $\phi_{1}$ and $\phi_{2}$
denote the upper turning points of the curves, which are the boundaries
related to the two highlighted regions I and II. (c) Cavity intensity
$\text{\ensuremath{\beta}}$ versus $\delta/\Gamma$ with $\phi=0$
(the left panel) and $\phi=-0.008\pi$ (the right panel). (d) The
sensitivity factor $\eta$ defined by Eq. (\ref{eq:SensitivityF})
plotted against $\phi$ and $\delta$. (e) $\eta$ versus $\phi$
for $\delta/\Gamma=0$. (f) $\eta$ versus $\delta/\Gamma$ with $\phi=0,\text{ }-0.008\pi$.
Other parameters are: $\Gamma/2\pi=100\text{ MHz}$, $\omega_{m}/2\pi=10$
kHz, $\kappa/\Gamma=2\times10^{-3}$, $P_{\text{in}}=8.06$ mW with
$\lambda_{d}=1550$ nm.}
\end{figure}

Since the weak single-photon optomechanical coupling $g/\omega_{m}$
is typically on the order ranging from $10^{-7}$ to $10^{-4}$ \citep{Aspelmeyer2013,Arcizet2006,Groeblacher2009,Groeblacher2009a,Kleckner2011,Rocheleau2010},
the optomechanical interaction induced Kerr nonlinearity $\chi=g^{2}/\omega_{m}$
can only be measured with a reasonably strong laser driving. Its effect
is reflected in the response of the cavity intensity $\beta$ of the
cavity mode $\hat{a}_{1}$. For the standard anti-PT symmetric system
with $\phi=0$ and $\kappa=0$, the response to $g$ at $\delta=0$
can be captured by the functional dependence $\beta\propto g^{-4/3}$
for $A=B=0$, according to which the sensitivity to $g$ is given
by $|\delta\beta/\delta g|\propto g^{-7/3}$. Thus, a triple decrease
in $g$ (or approximately a tenfold decrease in $\chi$) scales up
the peak value of $\beta$ by a factor of $\sim4$.

For $\phi\neq0$, the response of $\beta$ to $g$ can be examined
by solving Eq. (\ref{eq:beta_cubic}), which may have one or two stable
solutions. Here, $\beta$ is plotted against the detuning $\delta$
and the phase deviation $\phi$ in Fig. \ref{fig:4}(a) for the typical
set of parameters: $g/2\pi=3$ Hz, $\Gamma/2\pi=100$ MHz, $\kappa/\Gamma=2\times10^{-3}$,
and $P_{\text{in}}\approx8$ mW  with $\lambda_{d}=2\pi c/\omega_{d}=1550$
nm \citep{Shi2013,Dayan2008}. As expected, the bistability occurs
only for negative $\phi$. The asymmetry in the diagram with respect
to $\phi$ arises from the fact that the stability condition Eq. (\ref{eq:bistable condition})
does depend on the sign of the coherent coupling strength $\frac{\Gamma}{2}\text{sin\ensuremath{\phi}}$.
Note that for weak cavity driving, the detuning and the weak coherent
coupling associated with the bistable region, which is encircled by
the white dashed line, are limited to $|\delta/\Gamma|<0.14$ and
$-0.012\Gamma<\frac{\Gamma}{2}\text{sin\ensuremath{\phi}}<-0.003\Gamma$,
respectively. We emphasize that the bistability does not exist for
$\phi=0$, corresponding to the case of the completely dissipative
coupling regime.

For studying the sensitivity to $g$, we look into the modified bistability
behavior of $\beta$ by comparing its responses to two different optomechanical
coupling strengths $g=g_{1}=2\pi\times1$ Hz and $g=g_{2}=2\pi\times3$
Hz. As shown in Fig. \ref{fig:4}(b), we plot the cavity intensity
$\beta$ as a function of the phase deviation $\phi$ with $\delta=0$
for $g=g_{1}$ and $g=g_{2}$, respectively. In both cases, $\beta$
has two stable branches (solid) and one unstable branch (dashed).
Recalling that the perfect anti-PT symmetric system relies on the
fulfillment of the condition $\phi=0$, here when $\phi$ is slowly
swept backward, the cavity intensity $\beta$ responses to the weak
nonlinearity $\chi\propto g^{2}$ in a counter-intuitive way: the
system shows bistability for both the optomechanical coupling strengths
$g=g_{1}=2\pi\times1$ Hz and $g=g_{2}=2\pi\times3$ Hz, but the larger
$g$ (or the nonlinearity $\chi$) is, the smaller peak value of the
cavity intensity $\beta$ has. For comparison with the anti-PT symmetric
model, we show in Fig. \ref{fig:4}(c) the cavity intensity $\beta$
versus $\delta$ for $\phi=0$ (the left panel) and $\phi=-0.008\pi$
(the right panel), corresponding to the monostable and bistable regimes,
respectively. In both cases, $\beta$ shows flat central peak around
$\delta=0$, benefiting from the VIC induced linewidth suppression.
But for $\phi=-0.008\pi$, $\beta$ has a peak value about four (two)
times greater than that for $\phi=0$ with $g=g_{1}$ ($g=g_{2}$).
It proves that the optical bistability enables a stronger response
of $\beta$ to $g$. 

Considering the sensing of the OMIN $\chi$, there are generally two
working regions {[}as highlighted in Fig. \ref{fig:4}(b){]}, corresponding
to $\phi\in(\phi_{1},0)$ (region I) and $\phi\in(\phi_{2},\phi_{1})$
(region II), respectively. Here $\phi_{1}$ and $\phi_{2}$ are the
phase deviations corresponding to the turning points of the curves,
where the upper branch starts from and the middle unstable branch
ends. To be clear, we first define the sensitivity $\eta(\phi,\delta)$
as the ratio of the cavity intensities $\beta(g_{1})$ and $\beta(g_{2})$
at the \textit{upper branches} with respect to the two optomechanical
coupling strengths $g_{1}/2\pi=1\text{ Hz}$ and $g_{2}/2\pi=3\text{ Hz}$,
namely
\begin{eqnarray}
\eta(\phi,\delta) & = & \frac{\beta(g_{1})}{\beta(g_{2})},\label{eq:SensitivityF}
\end{eqnarray}
which is appropriate for region I. While for region II, $\eta^{-1}$
is used as the sensitivity instead. For an overview, we show $\eta(\phi,\delta)$
versus $\phi$ and $\delta$ in Fig. \ref{fig:4}(d), where the red
region corresponds to the enhanced sensitivity with the maximum $\eta_{\text{max}}\sim7.5$
for a triple decrease in $g$ (i.e. $g_{1}/g_{2}=1/3$). In region
I, for both coupling strengths the system displays monostability near
$\phi=0$ and optical bistability near $\phi=\phi_{1}$. The sensitivity
$\eta(\phi,0)$ at $\delta=0$ is gradually increased for a backward
sweep of the phase deviation from $\phi=0.003\pi$ to $\phi=\phi_{1}\approx-0.008\pi$,
as shown in Fig. \ref{fig:4}(e). It achieves the maximum $\eta(\phi_{1},0)\sim7.3$,
which is about twice that with $\phi=0$. Moreover, a slight detuning
allows for a better sensitivity for $\phi\approx-0.008\pi$, as manifested
by the $\delta$ dependence of $\eta$ shown in Fig. \ref{fig:4}(f).
The optimal $\eta(\phi=-0.008\pi,\delta)$ turns up at $\delta_{\text{opt}}\approx-0.025\Gamma$,
and is resistant to a detuning deviation $\sim0.03\Gamma$ from the
optimal detuning $\delta_{\text{opt}}$. 
\begin{flushleft}
\begin{figure}
\includegraphics[width=1\columnwidth]{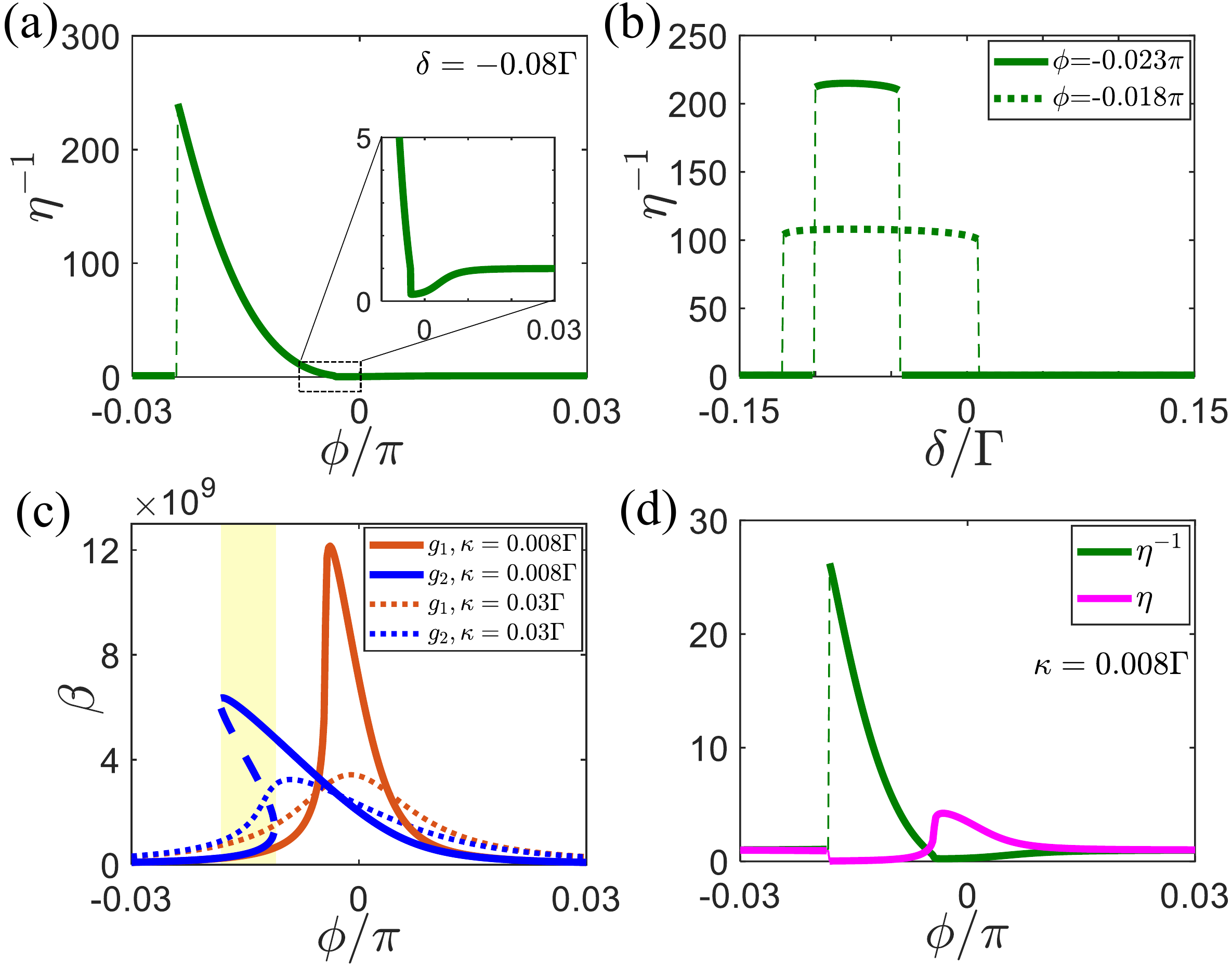}

\caption{\label{fig:5}(Color online) (a) The sensitivity $\eta^{-1}$ plotted
against $\phi$ at $\delta=-0.08\Gamma$; (b) $\eta^{-1}$ plotted
against $\delta$ at $\phi=-0.018\pi$, $-0.023\pi$; (c) Cavity intensity
$\text{\ensuremath{\beta}}$ on resonance $\delta=0$ versus $\phi$
for optomechanical coupling strengths $g_{1}/2\pi=1$ Hz and $g_{2}/2\pi=3$
Hz, where the cavity decay rates are set to $\kappa/\Gamma=0.008$
and 0.03. (d) The sensitivity factors $\eta$ and $\eta^{-1}$ versus
$\phi$ on resonance $\delta=0$ with $\kappa/\Gamma=0.008$. Other
parameters are the same as those in Fig. \ref{fig:4}.}
\end{figure}
\par\end{flushleft}

Next, we turn to region II in Fig. \ref{fig:4}(b), where $\beta$
becomes monostable for $g=g_{1}$ while maintains the bistability
feature for $g=g_{2}$. In this regime, the sensitivity factor $\eta$
approaches zero as $\phi$ sweeping backward and across $\phi_{1}$,
but the inverse of it (as the new sensitivity factor) $\eta^{-1}$
can be more than thirty times larger than the maximal $\eta$ at region
I. As shown in Fig. \ref{fig:5}(a), we plot $\eta^{-1}$ as a function
of $\phi$ with $\delta/\Gamma=-0.08$. Remarkably, $\eta^{-1}$ increases
from 0.3 to 240 in an almost monotonic manner when $\phi$ is swept
backward from $\phi=0$ to $\phi=\phi_{2}=-0.024\pi$, except a weak
dip at $\phi=-0.003\pi$ corresponding to the maximal $\beta(g_{1})$.
Furthermore, Fig. \ref{fig:5}(b) shows that the high sensitivity
(the maximum $\eta^{-1}$) remains flat within a certain frequency
range, which can be referred to as the bandwidth for the measurement
of the OMIN. Here the bandwidth is about $0.06\Gamma$ ($0.13\Gamma$)
for $\phi=-0.023\pi$ ($\phi=-0.018\pi$). Therefore, the coupled
cavity-waveguide system can act as a fluctuation-resistant and highly
sensitive sensor.

Finally, we consider the effect of the cavity decay by increasing
$\kappa$ and meanwhile keeping the cavity driving power invariant.
In Fig. \ref{fig:5}(c), we show $\beta$ as a function of $\phi$
with $\kappa/\Gamma=8\times10^{-3}$ and $\kappa/\Gamma=0.03$, respectively
\citep{Takashima2014}. For $\kappa/\Gamma=0.03$, the responses of
$\beta$ to $g_{1}/2\pi=1$ Hz and $g_{2}/2\pi=3$ Hz both show a
strong quenching and become monostable. The intensities decrease sharply
because a large cavity dissipation will inhibit the system from achieving
long-lived resonance and amplitude accumulation. Since the profiles
of the two dotted curves (with respect to $\kappa/\Gamma=0.03$) are
almost overlapped, the sensing scheme becomes invalid. While for a
weaker inherent decay $\kappa/\Gamma=8\times10^{-3}$, the behavior
observed at region II {[}shown in Fig. \ref{fig:4}(b){]} replays,
namely for $\phi\in(-0.018\pi,-0.011\pi)$, the response of $\beta$
to $g_{1}/2\pi=1$ Hz becomes monostable, while the response to $g_{2}/2\pi=3$
Hz remains bistable {[}highlighted in Fig. \ref{fig:5}(c){]}. Thus,
our model can still work as a high-performance sensor in this region
and the sensitivity factor $\eta^{-1}$ can still achieve a few tens.
In comparison, the sensitivity $\eta$ can only reach about 3 for
the anti-PT symmetric model with respect to $\phi=0$.

\section{Discussions on experimental realization}

To address the experimental feasibility, we consider an optomechanical
setup with a levitated nanosphere \citep{Neukirch2013,Neukirch2015,Hoang2016,Chang2010},
e.g. the silicon-vacancy (SiV) centers, coupling to the evanescent
field of a microsphere cavity with frequency $\omega_{c}$ \citep{Juan2016}.
The dielectric nanosphere with a nanodiamond structure contains $N$
two-level quantum emitters, which are driven by a bichromatic field
of frequencies symmetrically red- and blue-detuned from the atomic
transition frequency $\omega_{0}$, forming an optical trap for the
nanosphere at a distance $z$ from the cavity surface \citep{Juan2016}.
The emitter-cavity coupling $\Omega_{c}(\hat{z})$ then depends on
the mechanical displacement of the nanosphere along the $z$ axis,
with the mechanical eigenfrequency $\omega_{m}$. In the dispersive
interaction regime, where the emitter-cavity detuning $\Delta_{c}\equiv\omega_{c}-\omega_{0}$
is much larger than $\Omega_{c}$, and by considering that the induced
atomic level Stark shift is much less than the excited-state linewidth,
one can finally obtain the effective optomechanical interaction $\hat{H}_{\text{om}}=g\hat{a}^{\dagger}\hat{a}\hat{q}_{z}$
between the cavity field and the mechanical motion,  where \citep{Juan2016}
\begin{eqnarray}
g & = & -\sqrt{2}N(2p_{e}-1)\frac{\Omega_{c}^{2}(z)}{2\Delta_{c}}\gamma_{c}q_{\text{zpf}}
\end{eqnarray}
 is the single-photon coupling strength due to $N$ quantum emitters,
$p_{e}$ is the steady excited-state population, $\hat{q}_{z}$ is
the mechanical displacement operator along the $z$ axis from the
equilibrium position with $q_{\text{zpf}}$ the zero-point motion,
and $\gamma_{c}$ is the decay rate of the cavity\textquoteright s
evanescent field, see Ref. \citep{Juan2016} for the details.

The sensing device can then build on an optomechanical cavity-waveguide
system with two microcavities \citep{Farnesi2021,Chen2018,Takashima2014,Spillane2003},
one of which couples to an optically trapped nanosphere. For the typical
parameters \citep{Juan2016}: $p_{e}\approx3.2\times10^{-5}$, the
density of quantum emitters $\rho_{q}\approx1.4\text{ nm}^{-3}$,
the radius of the nanosphere $R=15$ nm, the mechanical frequency
$\omega_{m}/2\pi\sim10$ kHz, and $\gamma_{c}^{-1}=283$ nm, one obtains
the single-photon coupling strength $g/2\pi\sim$ 1 Hz (or $\chi/2\pi\sim100\text{ }\mu$Hz
) with the distance between the nanosphere and the cavity surface
being $z=783$ nm. Moreover, the inherent decay rate for the microcavity
is typically $\kappa/2\pi\sim10$ kHz, which is much less than the
waveguide induced coupling ($\sim$100 MHz) between the cavity modes
\citep{Dayan2008}. Therefore, the coupled cavity-waveguide device
can in principle be used for sensing weak perturbations on the set
of parameters with respect to the optomechanical interaction and the
induced Kerr nonlinearity. 

\section{Conclusion}

In summary, we have studied the sensitivity enhancement for measuring
the OMIN with two cavities coupled via a vacuum optical waveguide.
When the phase accumulation of light propagation from one cavity to
the other is precisely a multiple of $2\pi$, the two cavities are
dissipatively coupled. In this regime, the system can be anti-PT symmetric,
and its Hamiltonian shows a singular point around the cavity resonance
($\delta=0$), corresponding to the VIC induced zero linewidth. This
singularity has been exploited for sensing of a weak Kerr nonlinearity
\citep{Nair2021}, but can be fragile to a nonzero coherent coupling
and a finite cavity decay rate. In addition, the standard anti-PT
symmetric system shows bistability only at the broken anti-PT phase
with $|\delta/\Gamma|>1$ \citep{Nair2021b}. Although the standard
anti-PT symmetry cannot be held when a small phase deviation is introduced,
we find that both the VIC induced linewidth suppression and optical
bistability can be observed around the cavity resonance $\delta/\Gamma\sim0$.
The joint effect allows us to greatly enhance the sensitivity for
measuring the OMIN, which can reach a few hundreds for a triple reduction
of the single-photon optomechanical coupling strength. The high sensitivity
can be retained for a bandwidth of a few megahertz, and can remain
a few tens for a cavity decay rate of hundreds of kHz. While we consider
the setup consisting of an optically levitated nanodiamond coupled
to waveguide-coupled microspheres \citep{Juan2016}, the model can
be generalized to other integrated optomechanical cavity-waveguide
systems, such as hybrid cavity-magnonic systems \citep{Harder2018,ZareRameshti2018},
optomechanical crystal circuits \citep{Fang2016}, and microwave optomechanical
circuits \citep{Bernier2018}.

\section*{Acknowledgments}

H.W. acknowledges support from the National Natural Science Foundation
of China (NSFC) under Grants No. 11774058 and No. 12174058. Y.L. acknowledges
support from the NSFC under Grants No. 12074030 and No. 12274107.

\bibliographystyle{apsrev4-2}
\bibliography{Ref_antiPT}

\begin{thebibliography}{78}%
\makeatletter
\providecommand \@ifxundefined [1]{%
 \@ifx{#1\undefined}
}%
\providecommand \@ifnum [1]{%
 \ifnum #1\expandafter \@firstoftwo
 \else \expandafter \@secondoftwo
 \fi
}%
\providecommand \@ifx [1]{%
 \ifx #1\expandafter \@firstoftwo
 \else \expandafter \@secondoftwo
 \fi
}%
\providecommand \natexlab [1]{#1}%
\providecommand \enquote  [1]{``#1''}%
\providecommand \bibnamefont  [1]{#1}%
\providecommand \bibfnamefont [1]{#1}%
\providecommand \citenamefont [1]{#1}%
\providecommand \href@noop [0]{\@secondoftwo}%
\providecommand \href [0]{\begingroup \@sanitize@url \@href}%
\providecommand \@href[1]{\@@startlink{#1}\@@href}%
\providecommand \@@href[1]{\endgroup#1\@@endlink}%
\providecommand \@sanitize@url [0]{\catcode `\\12\catcode `\$12\catcode
  `\&12\catcode `\#12\catcode `\^12\catcode `\_12\catcode `\%12\relax}%
\providecommand \@@startlink[1]{}%
\providecommand \@@endlink[0]{}%
\providecommand \url  [0]{\begingroup\@sanitize@url \@url }%
\providecommand \@url [1]{\endgroup\@href {#1}{\urlprefix }}%
\providecommand \urlprefix  [0]{URL }%
\providecommand \Eprint [0]{\href }%
\providecommand \doibase [0]{https://doi.org/}%
\providecommand \selectlanguage [0]{\@gobble}%
\providecommand \bibinfo  [0]{\@secondoftwo}%
\providecommand \bibfield  [0]{\@secondoftwo}%
\providecommand \translation [1]{[#1]}%
\providecommand \BibitemOpen [0]{}%
\providecommand \bibitemStop [0]{}%
\providecommand \bibitemNoStop [0]{.\EOS\space}%
\providecommand \EOS [0]{\spacefactor3000\relax}%
\providecommand \BibitemShut  [1]{\csname bibitem#1\endcsname}%
\let\auto@bib@innerbib\@empty
\bibitem [{\citenamefont {Degen}\ \emph {et~al.}(2017)\citenamefont {Degen},
  \citenamefont {Reinhard},\ and\ \citenamefont {Cappellaro}}]{Degen2017}%
  \BibitemOpen
  \bibfield  {author} {\bibinfo {author} {\bibfnamefont {C.~L.}\ \bibnamefont
  {Degen}}, \bibinfo {author} {\bibfnamefont {F.}~\bibnamefont {Reinhard}},\
  and\ \bibinfo {author} {\bibfnamefont {P.}~\bibnamefont {Cappellaro}},\
  }\href {https://doi.org/10.1103/RevModPhys.89.035002} {\bibfield  {journal}
  {\bibinfo  {journal} {Rev. Mod. Phys.}\ }\textbf {\bibinfo {volume} {89}},\
  \bibinfo {pages} {035002} (\bibinfo {year} {2017})}\BibitemShut {NoStop}%
\bibitem [{\citenamefont {Brownnutt}\ \emph {et~al.}(2015)\citenamefont
  {Brownnutt}, \citenamefont {Kumph}, \citenamefont {Rabl},\ and\ \citenamefont
  {Blatt}}]{Brownnutt2015}%
  \BibitemOpen
  \bibfield  {author} {\bibinfo {author} {\bibfnamefont {M.}~\bibnamefont
  {Brownnutt}}, \bibinfo {author} {\bibfnamefont {M.}~\bibnamefont {Kumph}},
  \bibinfo {author} {\bibfnamefont {P.}~\bibnamefont {Rabl}},\ and\ \bibinfo
  {author} {\bibfnamefont {R.}~\bibnamefont {Blatt}},\ }\href
  {https://doi.org/10.1103/RevModPhys.87.1419} {\bibfield  {journal} {\bibinfo
  {journal} {Rev. Mod. Phys.}\ }\textbf {\bibinfo {volume} {87}},\ \bibinfo
  {pages} {1419} (\bibinfo {year} {2015})}\BibitemShut {NoStop}%
\bibitem [{\citenamefont {Jensen}\ \emph {et~al.}(2016)\citenamefont {Jensen},
  \citenamefont {Budvytyte}, \citenamefont {Thomas}, \citenamefont {Wang},
  \citenamefont {Fuchs}, \citenamefont {Balabas}, \citenamefont {Vasilakis},
  \citenamefont {Mosgaard}, \citenamefont {St{\ae}rkind}, \citenamefont
  {Muller}, \citenamefont {Heimburg}, \citenamefont {Olesen},\ and\
  \citenamefont {Polzik}}]{Jensen2016}%
  \BibitemOpen
  \bibfield  {author} {\bibinfo {author} {\bibfnamefont {K.}~\bibnamefont
  {Jensen}}, \bibinfo {author} {\bibfnamefont {R.}~\bibnamefont {Budvytyte}},
  \bibinfo {author} {\bibfnamefont {R.~A.}\ \bibnamefont {Thomas}}, \bibinfo
  {author} {\bibfnamefont {T.}~\bibnamefont {Wang}}, \bibinfo {author}
  {\bibfnamefont {A.~M.}\ \bibnamefont {Fuchs}}, \bibinfo {author}
  {\bibfnamefont {M.~V.}\ \bibnamefont {Balabas}}, \bibinfo {author}
  {\bibfnamefont {G.}~\bibnamefont {Vasilakis}}, \bibinfo {author}
  {\bibfnamefont {L.~D.}\ \bibnamefont {Mosgaard}}, \bibinfo {author}
  {\bibfnamefont {H.~C.}\ \bibnamefont {St{\ae}rkind}}, \bibinfo {author}
  {\bibfnamefont {J.~H.}\ \bibnamefont {Muller}}, \bibinfo {author}
  {\bibfnamefont {T.}~\bibnamefont {Heimburg}}, \bibinfo {author}
  {\bibfnamefont {S.~P.}\ \bibnamefont {Olesen}},\ and\ \bibinfo {author}
  {\bibfnamefont {E.~S.}\ \bibnamefont {Polzik}},\ }\href
  {https://doi.org/10.1038/srep29638} {\bibfield  {journal} {\bibinfo
  {journal} {Sci. Rep.}\ }\textbf {\bibinfo {volume} {6}},\ \bibinfo {pages}
  {29638} (\bibinfo {year} {2016})}\BibitemShut {NoStop}%
\bibitem [{\citenamefont {Schmitt}\ \emph {et~al.}(2017)\citenamefont
  {Schmitt}, \citenamefont {Gefen}, \citenamefont {St{\"{u}}rner},
  \citenamefont {Unden}, \citenamefont {Wolff}, \citenamefont {M{\"{u}}ller},
  \citenamefont {Scheuer}, \citenamefont {Naydenov}, \citenamefont {Markham},
  \citenamefont {Pezzagna}, \citenamefont {Meijer}, \citenamefont {Schwarz},
  \citenamefont {Plenio}, \citenamefont {Retzker}, \citenamefont {McGuinness},\
  and\ \citenamefont {Jelezko}}]{Schmitt2017}%
  \BibitemOpen
  \bibfield  {author} {\bibinfo {author} {\bibfnamefont {S.}~\bibnamefont
  {Schmitt}}, \bibinfo {author} {\bibfnamefont {T.}~\bibnamefont {Gefen}},
  \bibinfo {author} {\bibfnamefont {F.~M.}\ \bibnamefont {St{\"{u}}rner}},
  \bibinfo {author} {\bibfnamefont {T.}~\bibnamefont {Unden}}, \bibinfo
  {author} {\bibfnamefont {G.}~\bibnamefont {Wolff}}, \bibinfo {author}
  {\bibfnamefont {C.}~\bibnamefont {M{\"{u}}ller}}, \bibinfo {author}
  {\bibfnamefont {J.}~\bibnamefont {Scheuer}}, \bibinfo {author} {\bibfnamefont
  {B.}~\bibnamefont {Naydenov}}, \bibinfo {author} {\bibfnamefont
  {M.}~\bibnamefont {Markham}}, \bibinfo {author} {\bibfnamefont
  {S.}~\bibnamefont {Pezzagna}}, \bibinfo {author} {\bibfnamefont
  {J.}~\bibnamefont {Meijer}}, \bibinfo {author} {\bibfnamefont
  {I.}~\bibnamefont {Schwarz}}, \bibinfo {author} {\bibfnamefont
  {M.}~\bibnamefont {Plenio}}, \bibinfo {author} {\bibfnamefont
  {A.}~\bibnamefont {Retzker}}, \bibinfo {author} {\bibfnamefont {L.~P.}\
  \bibnamefont {McGuinness}},\ and\ \bibinfo {author} {\bibfnamefont
  {F.}~\bibnamefont {Jelezko}},\ }\href
  {https://doi.org/10.1126/science.aam5532} {\bibfield  {journal} {\bibinfo
  {journal} {Science}\ }\textbf {\bibinfo {volume} {356}},\ \bibinfo {pages}
  {832} (\bibinfo {year} {2017})}\BibitemShut {NoStop}%
\bibitem [{\citenamefont {Moser}\ \emph {et~al.}(2013)\citenamefont {Moser},
  \citenamefont {G{\"{u}}ttinger}, \citenamefont {Eichler}, \citenamefont
  {Esplandiu}, \citenamefont {Liu}, \citenamefont {Dykman},\ and\ \citenamefont
  {Bachtold}}]{Moser2013}%
  \BibitemOpen
  \bibfield  {author} {\bibinfo {author} {\bibfnamefont {J.}~\bibnamefont
  {Moser}}, \bibinfo {author} {\bibfnamefont {J.}~\bibnamefont
  {G{\"{u}}ttinger}}, \bibinfo {author} {\bibfnamefont {A.}~\bibnamefont
  {Eichler}}, \bibinfo {author} {\bibfnamefont {M.~J.}\ \bibnamefont
  {Esplandiu}}, \bibinfo {author} {\bibfnamefont {D.~E.}\ \bibnamefont {Liu}},
  \bibinfo {author} {\bibfnamefont {M.~I.}\ \bibnamefont {Dykman}},\ and\
  \bibinfo {author} {\bibfnamefont {A.}~\bibnamefont {Bachtold}},\ }\href
  {https://doi.org/10.1038/nnano.2013.97} {\bibfield  {journal} {\bibinfo
  {journal} {Nature Nanotech.}\ }\textbf {\bibinfo {volume} {8}},\ \bibinfo
  {pages} {493} (\bibinfo {year} {2013})}\BibitemShut {NoStop}%
\bibitem [{\citenamefont {Krause}\ \emph {et~al.}(2012)\citenamefont {Krause},
  \citenamefont {Winger}, \citenamefont {Blasius}, \citenamefont {Lin},\ and\
  \citenamefont {Painter}}]{Krause2012}%
  \BibitemOpen
  \bibfield  {author} {\bibinfo {author} {\bibfnamefont {A.~G.}\ \bibnamefont
  {Krause}}, \bibinfo {author} {\bibfnamefont {M.}~\bibnamefont {Winger}},
  \bibinfo {author} {\bibfnamefont {T.~D.}\ \bibnamefont {Blasius}}, \bibinfo
  {author} {\bibfnamefont {Q.}~\bibnamefont {Lin}},\ and\ \bibinfo {author}
  {\bibfnamefont {O.}~\bibnamefont {Painter}},\ }\href
  {https://doi.org/10.1038/nphoton.2012.245} {\bibfield  {journal} {\bibinfo
  {journal} {Nat. Photonics}\ }\textbf {\bibinfo {volume} {6}},\ \bibinfo
  {pages} {768} (\bibinfo {year} {2012})}\BibitemShut {NoStop}%
\bibitem [{\citenamefont {Gil-Santos}\ \emph {et~al.}(2010)\citenamefont
  {Gil-Santos}, \citenamefont {Ramos}, \citenamefont {Mart{\'{i}}nez},
  \citenamefont {Fern{\'{a}}ndez-Reg{\'{u}}lez}, \citenamefont {Garc{\'{i}}a},
  \citenamefont {{San Paulo}}, \citenamefont {Calleja},\ and\ \citenamefont
  {Tamayo}}]{GilSantos2010}%
  \BibitemOpen
  \bibfield  {author} {\bibinfo {author} {\bibfnamefont {E.}~\bibnamefont
  {Gil-Santos}}, \bibinfo {author} {\bibfnamefont {D.}~\bibnamefont {Ramos}},
  \bibinfo {author} {\bibfnamefont {J.}~\bibnamefont {Mart{\'{i}}nez}},
  \bibinfo {author} {\bibfnamefont {M.}~\bibnamefont
  {Fern{\'{a}}ndez-Reg{\'{u}}lez}}, \bibinfo {author} {\bibfnamefont
  {R.}~\bibnamefont {Garc{\'{i}}a}}, \bibinfo {author} {\bibfnamefont
  {{\'{A}}.}~\bibnamefont {{San Paulo}}}, \bibinfo {author} {\bibfnamefont
  {M.}~\bibnamefont {Calleja}},\ and\ \bibinfo {author} {\bibfnamefont
  {J.}~\bibnamefont {Tamayo}},\ }\href {https://doi.org/10.1038/nnano.2010.151}
  {\bibfield  {journal} {\bibinfo  {journal} {Nature Nanotech.}\ }\textbf
  {\bibinfo {volume} {5}},\ \bibinfo {pages} {641} (\bibinfo {year}
  {2010})}\BibitemShut {NoStop}%
\bibitem [{\citenamefont {Wiersig}(2014)}]{Wiersig2014}%
  \BibitemOpen
  \bibfield  {author} {\bibinfo {author} {\bibfnamefont {J.}~\bibnamefont
  {Wiersig}},\ }\href {https://doi.org/10.1103/PhysRevLett.112.203901}
  {\bibfield  {journal} {\bibinfo  {journal} {Phys. Rev. Lett.}\ }\textbf
  {\bibinfo {volume} {112}},\ \bibinfo {pages} {203901} (\bibinfo {year}
  {2014})}\BibitemShut {NoStop}%
\bibitem [{\citenamefont {Chang}\ \emph {et~al.}(2014)\citenamefont {Chang},
  \citenamefont {Jiang}, \citenamefont {Hua}, \citenamefont {Yang},
  \citenamefont {Wen}, \citenamefont {Jiang}, \citenamefont {Li}, \citenamefont
  {Wang},\ and\ \citenamefont {Xiao}}]{Chang_NatPhonon2014a}%
  \BibitemOpen
  \bibfield  {author} {\bibinfo {author} {\bibfnamefont {L.}~\bibnamefont
  {Chang}}, \bibinfo {author} {\bibfnamefont {X.}~\bibnamefont {Jiang}},
  \bibinfo {author} {\bibfnamefont {S.}~\bibnamefont {Hua}}, \bibinfo {author}
  {\bibfnamefont {C.}~\bibnamefont {Yang}}, \bibinfo {author} {\bibfnamefont
  {J.}~\bibnamefont {Wen}}, \bibinfo {author} {\bibfnamefont {L.}~\bibnamefont
  {Jiang}}, \bibinfo {author} {\bibfnamefont {G.}~\bibnamefont {Li}}, \bibinfo
  {author} {\bibfnamefont {G.}~\bibnamefont {Wang}},\ and\ \bibinfo {author}
  {\bibfnamefont {M.}~\bibnamefont {Xiao}},\ }\href
  {https://doi.org/10.1038/nphoton.2014.133} {\bibfield  {journal} {\bibinfo
  {journal} {Nat. Photonics}\ }\textbf {\bibinfo {volume} {8}},\ \bibinfo
  {pages} {524} (\bibinfo {year} {2014})}\BibitemShut {NoStop}%
\bibitem [{\citenamefont {Xiao}\ \emph {et~al.}(2019)\citenamefont {Xiao},
  \citenamefont {Li}, \citenamefont {Kottos},\ and\ \citenamefont
  {Al{\`{u}}}}]{Xiao2019}%
  \BibitemOpen
  \bibfield  {author} {\bibinfo {author} {\bibfnamefont {Z.}~\bibnamefont
  {Xiao}}, \bibinfo {author} {\bibfnamefont {H.}~\bibnamefont {Li}}, \bibinfo
  {author} {\bibfnamefont {T.}~\bibnamefont {Kottos}},\ and\ \bibinfo {author}
  {\bibfnamefont {A.}~\bibnamefont {Al{\`{u}}}},\ }\href
  {https://doi.org/10.1103/PhysRevLett.123.213901} {\bibfield  {journal}
  {\bibinfo  {journal} {Phys. Rev. Lett.}\ }\textbf {\bibinfo {volume} {123}},\
  \bibinfo {pages} {213901} (\bibinfo {year} {2019})}\BibitemShut {NoStop}%
\bibitem [{\citenamefont {D{\'{o}}ra}\ \emph {et~al.}(2021)\citenamefont
  {D{\'{o}}ra}, \citenamefont {Sticlet},\ and\ \citenamefont
  {Moca}}]{Dora2021}%
  \BibitemOpen
  \bibfield  {author} {\bibinfo {author} {\bibfnamefont {B.}~\bibnamefont
  {D{\'{o}}ra}}, \bibinfo {author} {\bibfnamefont {D.}~\bibnamefont
  {Sticlet}},\ and\ \bibinfo {author} {\bibfnamefont {C.~P.}\ \bibnamefont
  {Moca}},\ }\href {https://doi.org/10.1103/PhysRevLett.128.146804} {\bibfield
  {journal} {\bibinfo  {journal} {Phys. Rev. Lett.}\ }\textbf {\bibinfo
  {volume} {128}},\ \bibinfo {pages} {146804} (\bibinfo {year}
  {2021})}\BibitemShut {NoStop}%
\bibitem [{\citenamefont {Katsantonis}\ \emph {et~al.}(2022)\citenamefont
  {Katsantonis}, \citenamefont {Droulias}, \citenamefont {Soukoulis},
  \citenamefont {Economou}, \citenamefont {Rakitzis},\ and\ \citenamefont
  {Kafesaki}}]{Katsantonis2022}%
  \BibitemOpen
  \bibfield  {author} {\bibinfo {author} {\bibfnamefont {I.}~\bibnamefont
  {Katsantonis}}, \bibinfo {author} {\bibfnamefont {S.}~\bibnamefont
  {Droulias}}, \bibinfo {author} {\bibfnamefont {C.~M.}\ \bibnamefont
  {Soukoulis}}, \bibinfo {author} {\bibfnamefont {E.~N.}\ \bibnamefont
  {Economou}}, \bibinfo {author} {\bibfnamefont {T.~P.}\ \bibnamefont
  {Rakitzis}},\ and\ \bibinfo {author} {\bibfnamefont {M.}~\bibnamefont
  {Kafesaki}},\ }\href {https://doi.org/10.1103/PhysRevB.105.174112} {\bibfield
   {journal} {\bibinfo  {journal} {Phys. Rev. B}\ }\textbf {\bibinfo {volume}
  {105}},\ \bibinfo {pages} {174112} (\bibinfo {year} {2022})}\BibitemShut
  {NoStop}%
\bibitem [{\citenamefont {Ge}\ and\ \citenamefont
  {T{\"{u}}reci}(2013)}]{Ge2013}%
  \BibitemOpen
  \bibfield  {author} {\bibinfo {author} {\bibfnamefont {L.}~\bibnamefont
  {Ge}}\ and\ \bibinfo {author} {\bibfnamefont {H.~E.}\ \bibnamefont
  {T{\"{u}}reci}},\ }\href {https://doi.org/10.1103/PhysRevA.88.053810}
  {\bibfield  {journal} {\bibinfo  {journal} {Phys. Rev. A}\ }\textbf {\bibinfo
  {volume} {88}},\ \bibinfo {pages} {053810} (\bibinfo {year}
  {2013})}\BibitemShut {NoStop}%
\bibitem [{\citenamefont {Wang}\ \emph {et~al.}(2022)\citenamefont {Wang},
  \citenamefont {Mukhopadhyay},\ and\ \citenamefont {Agarwal}}]{Wang2022}%
  \BibitemOpen
  \bibfield  {author} {\bibinfo {author} {\bibfnamefont {J.}~\bibnamefont
  {Wang}}, \bibinfo {author} {\bibfnamefont {D.}~\bibnamefont {Mukhopadhyay}},\
  and\ \bibinfo {author} {\bibfnamefont {G.~S.}\ \bibnamefont {Agarwal}},\
  }\href {https://doi.org/10.1103/PhysRevResearch.4.013131} {\bibfield
  {journal} {\bibinfo  {journal} {Phys. Rev. Research}\ }\textbf {\bibinfo
  {volume} {4}},\ \bibinfo {pages} {013131} (\bibinfo {year}
  {2022})}\BibitemShut {NoStop}%
\bibitem [{\citenamefont {Luo}\ \emph {et~al.}(2022)\citenamefont {Luo},
  \citenamefont {Zhang},\ and\ \citenamefont {Du}}]{Luo2022}%
  \BibitemOpen
  \bibfield  {author} {\bibinfo {author} {\bibfnamefont {X.-W.}\ \bibnamefont
  {Luo}}, \bibinfo {author} {\bibfnamefont {C.}~\bibnamefont {Zhang}},\ and\
  \bibinfo {author} {\bibfnamefont {S.}~\bibnamefont {Du}},\ }\href
  {https://doi.org/10.1103/PhysRevLett.128.173602} {\bibfield  {journal}
  {\bibinfo  {journal} {Phys. Rev. Lett.}\ }\textbf {\bibinfo {volume} {128}},\
  \bibinfo {pages} {173602} (\bibinfo {year} {2022})}\BibitemShut {NoStop}%
\bibitem [{\citenamefont {Park}\ \emph {et~al.}(2021)\citenamefont {Park},
  \citenamefont {Lee}, \citenamefont {Park}, \citenamefont {Shin},
  \citenamefont {Choi},\ and\ \citenamefont {Yoon}}]{Park2021}%
  \BibitemOpen
  \bibfield  {author} {\bibinfo {author} {\bibfnamefont {S.}~\bibnamefont
  {Park}}, \bibinfo {author} {\bibfnamefont {D.}~\bibnamefont {Lee}}, \bibinfo
  {author} {\bibfnamefont {K.}~\bibnamefont {Park}}, \bibinfo {author}
  {\bibfnamefont {H.}~\bibnamefont {Shin}}, \bibinfo {author} {\bibfnamefont
  {Y.}~\bibnamefont {Choi}},\ and\ \bibinfo {author} {\bibfnamefont {J.~W.}\
  \bibnamefont {Yoon}},\ }\href
  {https://doi.org/10.1103/PhysRevLett.127.083601} {\bibfield  {journal}
  {\bibinfo  {journal} {Phys. Rev. Lett.}\ }\textbf {\bibinfo {volume} {127}},\
  \bibinfo {pages} {83601} (\bibinfo {year} {2021})}\BibitemShut {NoStop}%
\bibitem [{\citenamefont {Zhang}\ \emph {et~al.}(2020)\citenamefont {Zhang},
  \citenamefont {Huang}, \citenamefont {Zhang}, \citenamefont {Li},
  \citenamefont {Qiu}, \citenamefont {Nori},\ and\ \citenamefont
  {Jing}}]{Zhang2020}%
  \BibitemOpen
  \bibfield  {author} {\bibinfo {author} {\bibfnamefont {H.}~\bibnamefont
  {Zhang}}, \bibinfo {author} {\bibfnamefont {R.}~\bibnamefont {Huang}},
  \bibinfo {author} {\bibfnamefont {S.~D.}\ \bibnamefont {Zhang}}, \bibinfo
  {author} {\bibfnamefont {Y.}~\bibnamefont {Li}}, \bibinfo {author}
  {\bibfnamefont {C.~W.}\ \bibnamefont {Qiu}}, \bibinfo {author} {\bibfnamefont
  {F.}~\bibnamefont {Nori}},\ and\ \bibinfo {author} {\bibfnamefont
  {H.}~\bibnamefont {Jing}},\ }\href
  {https://doi.org/10.1021/acs.nanolett.0c03119} {\bibfield  {journal}
  {\bibinfo  {journal} {Nano Lett.}\ }\textbf {\bibinfo {volume} {20}},\
  \bibinfo {pages} {7594} (\bibinfo {year} {2020})}\BibitemShut {NoStop}%
\bibitem [{\citenamefont {Nair}\ \emph
  {et~al.}(2021{\natexlab{a}})\citenamefont {Nair}, \citenamefont
  {Mukhopadhyay},\ and\ \citenamefont {Agarwal}}]{Nair2021}%
  \BibitemOpen
  \bibfield  {author} {\bibinfo {author} {\bibfnamefont {J.~M.}\ \bibnamefont
  {Nair}}, \bibinfo {author} {\bibfnamefont {D.}~\bibnamefont {Mukhopadhyay}},\
  and\ \bibinfo {author} {\bibfnamefont {G.~S.}\ \bibnamefont {Agarwal}},\
  }\href {https://doi.org/10.1103/PhysRevLett.126.180401} {\bibfield  {journal}
  {\bibinfo  {journal} {Phys. Rev. Lett.}\ }\textbf {\bibinfo {volume} {126}},\
  \bibinfo {pages} {180401} (\bibinfo {year} {2021}{\natexlab{a}})}\BibitemShut
  {NoStop}%
\bibitem [{\citenamefont {Chen}\ \emph {et~al.}(2017)\citenamefont {Chen},
  \citenamefont {{Kaya {\"{O}}zdemir}}, \citenamefont {Zhao}, \citenamefont
  {Wiersig},\ and\ \citenamefont {Yang}}]{Chen2017}%
  \BibitemOpen
  \bibfield  {author} {\bibinfo {author} {\bibfnamefont {W.}~\bibnamefont
  {Chen}}, \bibinfo {author} {\bibfnamefont {S.}~\bibnamefont {{Kaya
  {\"{O}}zdemir}}}, \bibinfo {author} {\bibfnamefont {G.}~\bibnamefont {Zhao}},
  \bibinfo {author} {\bibfnamefont {J.}~\bibnamefont {Wiersig}},\ and\ \bibinfo
  {author} {\bibfnamefont {L.}~\bibnamefont {Yang}},\ }\href
  {https://doi.org/10.1038/nature23281} {\bibfield  {journal} {\bibinfo
  {journal} {Nature (London)}\ }\textbf {\bibinfo {volume} {548}},\ \bibinfo
  {pages} {192} (\bibinfo {year} {2017})}\BibitemShut {NoStop}%
\bibitem [{\citenamefont {Hodaei}\ \emph {et~al.}(2017)\citenamefont {Hodaei},
  \citenamefont {Hassan}, \citenamefont {Wittek}, \citenamefont
  {Garcia-Gracia}, \citenamefont {El-Ganainy}, \citenamefont
  {Christodoulides},\ and\ \citenamefont {Khajavikhan}}]{Hodaei2017}%
  \BibitemOpen
  \bibfield  {author} {\bibinfo {author} {\bibfnamefont {H.}~\bibnamefont
  {Hodaei}}, \bibinfo {author} {\bibfnamefont {A.~U.}\ \bibnamefont {Hassan}},
  \bibinfo {author} {\bibfnamefont {S.}~\bibnamefont {Wittek}}, \bibinfo
  {author} {\bibfnamefont {H.}~\bibnamefont {Garcia-Gracia}}, \bibinfo {author}
  {\bibfnamefont {R.}~\bibnamefont {El-Ganainy}}, \bibinfo {author}
  {\bibfnamefont {D.~N.}\ \bibnamefont {Christodoulides}},\ and\ \bibinfo
  {author} {\bibfnamefont {M.}~\bibnamefont {Khajavikhan}},\ }\href
  {https://doi.org/10.1038/nature23280} {\bibfield  {journal} {\bibinfo
  {journal} {Nature (London)}\ }\textbf {\bibinfo {volume} {548}},\ \bibinfo
  {pages} {187} (\bibinfo {year} {2017})}\BibitemShut {NoStop}%
\bibitem [{\citenamefont {El-Ganainy}\ \emph {et~al.}(2018)\citenamefont
  {El-Ganainy}, \citenamefont {Makris}, \citenamefont {Khajavikhan},
  \citenamefont {Musslimani}, \citenamefont {Rotter},\ and\ \citenamefont
  {Christodoulides}}]{El-Ganainy2018}%
  \BibitemOpen
  \bibfield  {author} {\bibinfo {author} {\bibfnamefont {R.}~\bibnamefont
  {El-Ganainy}}, \bibinfo {author} {\bibfnamefont {K.~G.}\ \bibnamefont
  {Makris}}, \bibinfo {author} {\bibfnamefont {M.}~\bibnamefont {Khajavikhan}},
  \bibinfo {author} {\bibfnamefont {Z.~H.}\ \bibnamefont {Musslimani}},
  \bibinfo {author} {\bibfnamefont {S.}~\bibnamefont {Rotter}},\ and\ \bibinfo
  {author} {\bibfnamefont {D.~N.}\ \bibnamefont {Christodoulides}},\ }\href
  {https://doi.org/10.1038/nphys4323} {\bibfield  {journal} {\bibinfo
  {journal} {Nat. Phys.}\ }\textbf {\bibinfo {volume} {14}},\ \bibinfo {pages}
  {11} (\bibinfo {year} {2018})}\BibitemShut {NoStop}%
\bibitem [{\citenamefont {Miri}\ and\ \citenamefont
  {Al{\`{u}}}(2019)}]{Miri2019}%
  \BibitemOpen
  \bibfield  {author} {\bibinfo {author} {\bibfnamefont {M.~A.}\ \bibnamefont
  {Miri}}\ and\ \bibinfo {author} {\bibfnamefont {A.}~\bibnamefont
  {Al{\`{u}}}},\ }\href {https://doi.org/10.1126/science.aar7709} {\bibfield
  {journal} {\bibinfo  {journal} {Science}\ }\textbf {\bibinfo {volume}
  {363}},\ \bibinfo {pages} {7709} (\bibinfo {year} {2019})}\BibitemShut
  {NoStop}%
\bibitem [{\citenamefont {Li}\ \emph {et~al.}(2020)\citenamefont {Li},
  \citenamefont {Mekawy}, \citenamefont {Krasnok},\ and\ \citenamefont
  {Al{\`{u}}}}]{Li2020}%
  \BibitemOpen
  \bibfield  {author} {\bibinfo {author} {\bibfnamefont {H.}~\bibnamefont
  {Li}}, \bibinfo {author} {\bibfnamefont {A.}~\bibnamefont {Mekawy}}, \bibinfo
  {author} {\bibfnamefont {A.}~\bibnamefont {Krasnok}},\ and\ \bibinfo {author}
  {\bibfnamefont {A.}~\bibnamefont {Al{\`{u}}}},\ }\href
  {https://doi.org/10.1103/PhysRevLett.124.193901} {\bibfield  {journal}
  {\bibinfo  {journal} {Phys. Rev. Lett.}\ }\textbf {\bibinfo {volume} {124}},\
  \bibinfo {pages} {193901} (\bibinfo {year} {2020})}\BibitemShut {NoStop}%
\bibitem [{\citenamefont {Yang}\ \emph {et~al.}(2020)\citenamefont {Yang},
  \citenamefont {Wang}, \citenamefont {Rao}, \citenamefont {Gui}, \citenamefont
  {Yao}, \citenamefont {Lu},\ and\ \citenamefont {Hu}}]{Yang2020}%
  \BibitemOpen
  \bibfield  {author} {\bibinfo {author} {\bibfnamefont {Y.}~\bibnamefont
  {Yang}}, \bibinfo {author} {\bibfnamefont {Y.~P.}\ \bibnamefont {Wang}},
  \bibinfo {author} {\bibfnamefont {J.~W.}\ \bibnamefont {Rao}}, \bibinfo
  {author} {\bibfnamefont {Y.~S.}\ \bibnamefont {Gui}}, \bibinfo {author}
  {\bibfnamefont {B.~M.}\ \bibnamefont {Yao}}, \bibinfo {author} {\bibfnamefont
  {W.}~\bibnamefont {Lu}},\ and\ \bibinfo {author} {\bibfnamefont {C.~M.}\
  \bibnamefont {Hu}},\ }\href {https://doi.org/10.1103/PhysRevLett.125.147202}
  {\bibfield  {journal} {\bibinfo  {journal} {Phys. Rev. Lett.}\ }\textbf
  {\bibinfo {volume} {125}},\ \bibinfo {pages} {147202} (\bibinfo {year}
  {2020})}\BibitemShut {NoStop}%
\bibitem [{\citenamefont {Peng}\ \emph {et~al.}(2016)\citenamefont {Peng},
  \citenamefont {Cao}, \citenamefont {Shen}, \citenamefont {Qu}, \citenamefont
  {Wen}, \citenamefont {Jiang},\ and\ \citenamefont {Xiao}}]{Peng2016}%
  \BibitemOpen
  \bibfield  {author} {\bibinfo {author} {\bibfnamefont {P.}~\bibnamefont
  {Peng}}, \bibinfo {author} {\bibfnamefont {W.}~\bibnamefont {Cao}}, \bibinfo
  {author} {\bibfnamefont {C.}~\bibnamefont {Shen}}, \bibinfo {author}
  {\bibfnamefont {W.}~\bibnamefont {Qu}}, \bibinfo {author} {\bibfnamefont
  {J.}~\bibnamefont {Wen}}, \bibinfo {author} {\bibfnamefont {L.}~\bibnamefont
  {Jiang}},\ and\ \bibinfo {author} {\bibfnamefont {Y.}~\bibnamefont {Xiao}},\
  }\href {https://doi.org/10.1038/nphys3842} {\bibfield  {journal} {\bibinfo
  {journal} {Nat. Phys.}\ }\textbf {\bibinfo {volume} {12}},\ \bibinfo {pages}
  {1139} (\bibinfo {year} {2016})}\BibitemShut {NoStop}%
\bibitem [{\citenamefont {Yin}\ and\ \citenamefont {Zhang}(2013)}]{Yin2013}%
  \BibitemOpen
  \bibfield  {author} {\bibinfo {author} {\bibfnamefont {X.}~\bibnamefont
  {Yin}}\ and\ \bibinfo {author} {\bibfnamefont {X.}~\bibnamefont {Zhang}},\
  }\href {https://doi.org/10.1038/nmat3576} {\bibfield  {journal} {\bibinfo
  {journal} {Nat. Mater.}\ }\textbf {\bibinfo {volume} {12}},\ \bibinfo {pages}
  {175} (\bibinfo {year} {2013})}\BibitemShut {NoStop}%
\bibitem [{\citenamefont {Feng}\ \emph {et~al.}(2011)\citenamefont {Feng},
  \citenamefont {Ayache}, \citenamefont {Huang}, \citenamefont {Xu},
  \citenamefont {Lu}, \citenamefont {Chen}, \citenamefont {Fainman},\ and\
  \citenamefont {Scherer}}]{White2011}%
  \BibitemOpen
  \bibfield  {author} {\bibinfo {author} {\bibfnamefont {L.}~\bibnamefont
  {Feng}}, \bibinfo {author} {\bibfnamefont {M.}~\bibnamefont {Ayache}},
  \bibinfo {author} {\bibfnamefont {J.}~\bibnamefont {Huang}}, \bibinfo
  {author} {\bibfnamefont {Y.-L.}\ \bibnamefont {Xu}}, \bibinfo {author}
  {\bibfnamefont {M.-H.}\ \bibnamefont {Lu}}, \bibinfo {author} {\bibfnamefont
  {Y.-F.}\ \bibnamefont {Chen}}, \bibinfo {author} {\bibfnamefont
  {Y.}~\bibnamefont {Fainman}},\ and\ \bibinfo {author} {\bibfnamefont
  {A.}~\bibnamefont {Scherer}},\ }\href
  {https://doi.org/10.1126/science.1206038} {\bibfield  {journal} {\bibinfo
  {journal} {Science}\ }\textbf {\bibinfo {volume} {333}},\ \bibinfo {pages}
  {729} (\bibinfo {year} {2011})}\BibitemShut {NoStop}%
\bibitem [{\citenamefont {Feng}\ \emph {et~al.}(2014)\citenamefont {Feng},
  \citenamefont {Wong}, \citenamefont {Ma}, \citenamefont {Wang},\ and\
  \citenamefont {Zhang}}]{Feng2014}%
  \BibitemOpen
  \bibfield  {author} {\bibinfo {author} {\bibfnamefont {L.}~\bibnamefont
  {Feng}}, \bibinfo {author} {\bibfnamefont {Z.~J.}\ \bibnamefont {Wong}},
  \bibinfo {author} {\bibfnamefont {R.~M.}\ \bibnamefont {Ma}}, \bibinfo
  {author} {\bibfnamefont {Y.}~\bibnamefont {Wang}},\ and\ \bibinfo {author}
  {\bibfnamefont {X.}~\bibnamefont {Zhang}},\ }\href
  {https://doi.org/10.1126/science.1258479} {\bibfield  {journal} {\bibinfo
  {journal} {Science}\ }\textbf {\bibinfo {volume} {346}},\ \bibinfo {pages}
  {972} (\bibinfo {year} {2014})}\BibitemShut {NoStop}%
\bibitem [{\citenamefont {Hodaei}\ \emph {et~al.}(2014)\citenamefont {Hodaei},
  \citenamefont {Miri}, \citenamefont {Heinrich}, \citenamefont
  {Christodoulides},\ and\ \citenamefont
  {Khajavikhan}}]{HosseinHodaeiMohammad-AliMiri2014}%
  \BibitemOpen
  \bibfield  {author} {\bibinfo {author} {\bibfnamefont {H.}~\bibnamefont
  {Hodaei}}, \bibinfo {author} {\bibfnamefont {M.~A.}\ \bibnamefont {Miri}},
  \bibinfo {author} {\bibfnamefont {M.}~\bibnamefont {Heinrich}}, \bibinfo
  {author} {\bibfnamefont {D.~N.}\ \bibnamefont {Christodoulides}},\ and\
  \bibinfo {author} {\bibfnamefont {M.}~\bibnamefont {Khajavikhan}},\ }\href
  {https://doi.org/10.1126/science.1258480} {\bibfield  {journal} {\bibinfo
  {journal} {Science}\ }\textbf {\bibinfo {volume} {346}},\ \bibinfo {pages}
  {975} (\bibinfo {year} {2014})}\BibitemShut {NoStop}%
\bibitem [{\citenamefont {Yoon}\ \emph {et~al.}(2018)\citenamefont {Yoon},
  \citenamefont {Choi}, \citenamefont {Hahn}, \citenamefont {Kim},
  \citenamefont {Song}, \citenamefont {Yang}, \citenamefont {Lee},
  \citenamefont {Kim}, \citenamefont {Lee}, \citenamefont {Shin}, \citenamefont
  {Lee},\ and\ \citenamefont {Berini}}]{Yoon2018}%
  \BibitemOpen
  \bibfield  {author} {\bibinfo {author} {\bibfnamefont {J.~W.}\ \bibnamefont
  {Yoon}}, \bibinfo {author} {\bibfnamefont {Y.}~\bibnamefont {Choi}}, \bibinfo
  {author} {\bibfnamefont {C.}~\bibnamefont {Hahn}}, \bibinfo {author}
  {\bibfnamefont {G.}~\bibnamefont {Kim}}, \bibinfo {author} {\bibfnamefont
  {S.~H.}\ \bibnamefont {Song}}, \bibinfo {author} {\bibfnamefont {K.-Y.}\
  \bibnamefont {Yang}}, \bibinfo {author} {\bibfnamefont {J.~Y.}\ \bibnamefont
  {Lee}}, \bibinfo {author} {\bibfnamefont {Y.}~\bibnamefont {Kim}}, \bibinfo
  {author} {\bibfnamefont {C.~S.}\ \bibnamefont {Lee}}, \bibinfo {author}
  {\bibfnamefont {J.~K.}\ \bibnamefont {Shin}}, \bibinfo {author}
  {\bibfnamefont {H.-S.}\ \bibnamefont {Lee}},\ and\ \bibinfo {author}
  {\bibfnamefont {P.}~\bibnamefont {Berini}},\ }\href
  {https://doi.org/10.1038/s41586-018-0523-2} {\bibfield  {journal} {\bibinfo
  {journal} {Nature (London)}\ }\textbf {\bibinfo {volume} {562}},\ \bibinfo
  {pages} {86} (\bibinfo {year} {2018})}\BibitemShut {NoStop}%
\bibitem [{\citenamefont {Zhang}\ \emph {et~al.}(2019)\citenamefont {Zhang},
  \citenamefont {Jiang},\ and\ \citenamefont {Chan}}]{Zhang2019}%
  \BibitemOpen
  \bibfield  {author} {\bibinfo {author} {\bibfnamefont {X.~L.}\ \bibnamefont
  {Zhang}}, \bibinfo {author} {\bibfnamefont {T.}~\bibnamefont {Jiang}},\ and\
  \bibinfo {author} {\bibfnamefont {C.~T.}\ \bibnamefont {Chan}},\ }\href
  {https://doi.org/10.1038/s41377-019-0200-8} {\bibfield  {journal} {\bibinfo
  {journal} {Light: Sci. Appl.}\ }\textbf {\bibinfo {volume} {8}},\ \bibinfo
  {pages} {88} (\bibinfo {year} {2019})}\BibitemShut {NoStop}%
\bibitem [{\citenamefont {Chen}\ and\ \citenamefont {Lee}(2018)}]{Chen2018}%
  \BibitemOpen
  \bibfield  {author} {\bibinfo {author} {\bibfnamefont {T.~C.}\ \bibnamefont
  {Chen}}\ and\ \bibinfo {author} {\bibfnamefont {M.~C.~M.}\ \bibnamefont
  {Lee}},\ }\href {https://doi.org/10.1109/JLT.2018.2832047} {\bibfield
  {journal} {\bibinfo  {journal} {J. Lightw. Technol.}\ }\textbf {\bibinfo
  {volume} {36}},\ \bibinfo {pages} {2966} (\bibinfo {year}
  {2018})}\BibitemShut {NoStop}%
\bibitem [{\citenamefont {Dong}\ \emph {et~al.}(2019)\citenamefont {Dong},
  \citenamefont {Li}, \citenamefont {Yang}, \citenamefont {Qiu},\ and\
  \citenamefont {Ho}}]{Dong2019}%
  \BibitemOpen
  \bibfield  {author} {\bibinfo {author} {\bibfnamefont {Z.}~\bibnamefont
  {Dong}}, \bibinfo {author} {\bibfnamefont {Z.}~\bibnamefont {Li}}, \bibinfo
  {author} {\bibfnamefont {F.}~\bibnamefont {Yang}}, \bibinfo {author}
  {\bibfnamefont {C.-W.}\ \bibnamefont {Qiu}},\ and\ \bibinfo {author}
  {\bibfnamefont {J.~S.}\ \bibnamefont {Ho}},\ }\href
  {https://doi.org/10.1038/s41928-019-0284-4} {\bibfield  {journal} {\bibinfo
  {journal} {Nat. Electron.}\ }\textbf {\bibinfo {volume} {2}},\ \bibinfo
  {pages} {335} (\bibinfo {year} {2019})}\BibitemShut {NoStop}%
\bibitem [{\citenamefont {Choi}\ \emph {et~al.}(2018)\citenamefont {Choi},
  \citenamefont {Hahn}, \citenamefont {Yoon},\ and\ \citenamefont
  {Song}}]{Choi2018}%
  \BibitemOpen
  \bibfield  {author} {\bibinfo {author} {\bibfnamefont {Y.}~\bibnamefont
  {Choi}}, \bibinfo {author} {\bibfnamefont {C.}~\bibnamefont {Hahn}}, \bibinfo
  {author} {\bibfnamefont {J.~W.}\ \bibnamefont {Yoon}},\ and\ \bibinfo
  {author} {\bibfnamefont {S.~H.}\ \bibnamefont {Song}},\ }\href
  {https://doi.org/10.1038/s41467-018-04690-y} {\bibfield  {journal} {\bibinfo
  {journal} {Nat. Commun.}\ }\textbf {\bibinfo {volume} {9}},\ \bibinfo {pages}
  {2182} (\bibinfo {year} {2018})}\BibitemShut {NoStop}%
\bibitem [{\citenamefont {Paspalakis}\ and\ \citenamefont
  {Knight}(1998)}]{Paspalakis1998}%
  \BibitemOpen
  \bibfield  {author} {\bibinfo {author} {\bibfnamefont {E.}~\bibnamefont
  {Paspalakis}}\ and\ \bibinfo {author} {\bibfnamefont {P.~L.}\ \bibnamefont
  {Knight}},\ }\href {https://doi.org/10.1103/PhysRevLett.81.293} {\bibfield
  {journal} {\bibinfo  {journal} {Phys. Rev. Lett.}\ }\textbf {\bibinfo
  {volume} {81}},\ \bibinfo {pages} {293} (\bibinfo {year} {1998})}\BibitemShut
  {NoStop}%
\bibitem [{\citenamefont {Keitel}(1999)}]{Keitel1999}%
  \BibitemOpen
  \bibfield  {author} {\bibinfo {author} {\bibfnamefont {C.~H.}\ \bibnamefont
  {Keitel}},\ }\href {https://doi.org/10.1103/PhysRevLett.83.1307} {\bibfield
  {journal} {\bibinfo  {journal} {Phys. Rev. Lett.}\ }\textbf {\bibinfo
  {volume} {83}},\ \bibinfo {pages} {1307} (\bibinfo {year}
  {1999})}\BibitemShut {NoStop}%
\bibitem [{\citenamefont {Agarwal}(2000)}]{Agarwal2000}%
  \BibitemOpen
  \bibfield  {author} {\bibinfo {author} {\bibfnamefont {G.~S.}\ \bibnamefont
  {Agarwal}},\ }\href {https://doi.org/10.1103/PhysRevLett.84.5500} {\bibfield
  {journal} {\bibinfo  {journal} {Phys. Rev. Lett.}\ }\textbf {\bibinfo
  {volume} {84}},\ \bibinfo {pages} {5500} (\bibinfo {year}
  {2000})}\BibitemShut {NoStop}%
\bibitem [{\citenamefont {Scully}(2010)}]{Scully2010}%
  \BibitemOpen
  \bibfield  {author} {\bibinfo {author} {\bibfnamefont {M.~O.}\ \bibnamefont
  {Scully}},\ }\href {https://doi.org/10.1103/PhysRevLett.104.207701}
  {\bibfield  {journal} {\bibinfo  {journal} {Phys. Rev. Lett.}\ }\textbf
  {\bibinfo {volume} {104}},\ \bibinfo {pages} {207701} (\bibinfo {year}
  {2010})}\BibitemShut {NoStop}%
\bibitem [{\citenamefont {Heeg}\ \emph {et~al.}(2013)\citenamefont {Heeg},
  \citenamefont {Wille}, \citenamefont {Schlage}, \citenamefont {Guryeva},
  \citenamefont {Schumacher}, \citenamefont {Uschmann}, \citenamefont
  {Schulze}, \citenamefont {Marx}, \citenamefont {K{\"{a}}mpfer}, \citenamefont
  {Paulus}, \citenamefont {R{\"{o}}hlsberger},\ and\ \citenamefont
  {Evers}}]{Heeg2013}%
  \BibitemOpen
  \bibfield  {author} {\bibinfo {author} {\bibfnamefont {K.~P.}\ \bibnamefont
  {Heeg}}, \bibinfo {author} {\bibfnamefont {H.~C.}\ \bibnamefont {Wille}},
  \bibinfo {author} {\bibfnamefont {K.}~\bibnamefont {Schlage}}, \bibinfo
  {author} {\bibfnamefont {T.}~\bibnamefont {Guryeva}}, \bibinfo {author}
  {\bibfnamefont {D.}~\bibnamefont {Schumacher}}, \bibinfo {author}
  {\bibfnamefont {I.}~\bibnamefont {Uschmann}}, \bibinfo {author}
  {\bibfnamefont {K.~S.}\ \bibnamefont {Schulze}}, \bibinfo {author}
  {\bibfnamefont {B.}~\bibnamefont {Marx}}, \bibinfo {author} {\bibfnamefont
  {T.}~\bibnamefont {K{\"{a}}mpfer}}, \bibinfo {author} {\bibfnamefont {G.~G.}\
  \bibnamefont {Paulus}}, \bibinfo {author} {\bibfnamefont {R.}~\bibnamefont
  {R{\"{o}}hlsberger}},\ and\ \bibinfo {author} {\bibfnamefont
  {J.}~\bibnamefont {Evers}},\ }\href
  {https://doi.org/10.1103/PhysRevLett.111.073601} {\bibfield  {journal}
  {\bibinfo  {journal} {Phys. Rev. Lett.}\ }\textbf {\bibinfo {volume} {111}},\
  \bibinfo {pages} {073601} (\bibinfo {year} {2013})}\BibitemShut {NoStop}%
\bibitem [{\citenamefont {Jha}\ \emph {et~al.}(2015)\citenamefont {Jha},
  \citenamefont {Ni}, \citenamefont {Wu}, \citenamefont {Wang},\ and\
  \citenamefont {Zhang}}]{Jha2015}%
  \BibitemOpen
  \bibfield  {author} {\bibinfo {author} {\bibfnamefont {P.~K.}\ \bibnamefont
  {Jha}}, \bibinfo {author} {\bibfnamefont {X.}~\bibnamefont {Ni}}, \bibinfo
  {author} {\bibfnamefont {C.}~\bibnamefont {Wu}}, \bibinfo {author}
  {\bibfnamefont {Y.}~\bibnamefont {Wang}},\ and\ \bibinfo {author}
  {\bibfnamefont {X.}~\bibnamefont {Zhang}},\ }\href
  {https://doi.org/10.1103/PhysRevLett.115.025501} {\bibfield  {journal}
  {\bibinfo  {journal} {Phys. Rev. Lett.}\ }\textbf {\bibinfo {volume} {115}},\
  \bibinfo {pages} {025501} (\bibinfo {year} {2015})}\BibitemShut {NoStop}%
\bibitem [{\citenamefont {Hisatomi}\ \emph {et~al.}(2016)\citenamefont
  {Hisatomi}, \citenamefont {Osada}, \citenamefont {Tabuchi}, \citenamefont
  {Ishikawa}, \citenamefont {Noguchi}, \citenamefont {Yamazaki}, \citenamefont
  {Usami},\ and\ \citenamefont {Nakamura}}]{Hisatomi2016}%
  \BibitemOpen
  \bibfield  {author} {\bibinfo {author} {\bibfnamefont {R.}~\bibnamefont
  {Hisatomi}}, \bibinfo {author} {\bibfnamefont {A.}~\bibnamefont {Osada}},
  \bibinfo {author} {\bibfnamefont {Y.}~\bibnamefont {Tabuchi}}, \bibinfo
  {author} {\bibfnamefont {T.}~\bibnamefont {Ishikawa}}, \bibinfo {author}
  {\bibfnamefont {A.}~\bibnamefont {Noguchi}}, \bibinfo {author} {\bibfnamefont
  {R.}~\bibnamefont {Yamazaki}}, \bibinfo {author} {\bibfnamefont
  {K.}~\bibnamefont {Usami}},\ and\ \bibinfo {author} {\bibfnamefont
  {Y.}~\bibnamefont {Nakamura}},\ }\href
  {https://doi.org/10.1103/PhysRevB.93.174427} {\bibfield  {journal} {\bibinfo
  {journal} {Phys. Rev. B}\ }\textbf {\bibinfo {volume} {93}},\ \bibinfo
  {pages} {174427} (\bibinfo {year} {2016})}\BibitemShut {NoStop}%
\bibitem [{\citenamefont {Konotop}\ and\ \citenamefont
  {Zezyulin}(2018)}]{Konotop2018}%
  \BibitemOpen
  \bibfield  {author} {\bibinfo {author} {\bibfnamefont {V.~V.}\ \bibnamefont
  {Konotop}}\ and\ \bibinfo {author} {\bibfnamefont {D.~A.}\ \bibnamefont
  {Zezyulin}},\ }\href {https://doi.org/10.1103/PhysRevLett.120.123902}
  {\bibfield  {journal} {\bibinfo  {journal} {Phys. Rev. Lett.}\ }\textbf
  {\bibinfo {volume} {120}},\ \bibinfo {pages} {123902} (\bibinfo {year}
  {2018})},\ \Eprint {https://arxiv.org/abs/1802.07674} {1802.07674}
  \BibitemShut {NoStop}%
\bibitem [{\citenamefont {Mukhopadhyay}\ \emph {et~al.}(2022)\citenamefont
  {Mukhopadhyay}, \citenamefont {Nair},\ and\ \citenamefont
  {Agarwal}}]{Mukhopadhyay2022}%
  \BibitemOpen
  \bibfield  {author} {\bibinfo {author} {\bibfnamefont {D.}~\bibnamefont
  {Mukhopadhyay}}, \bibinfo {author} {\bibfnamefont {J.~M.}\ \bibnamefont
  {Nair}},\ and\ \bibinfo {author} {\bibfnamefont {G.~S.}\ \bibnamefont
  {Agarwal}},\ }\href {https://doi.org/10.1103/PhysRevB.105.064405} {\bibfield
  {journal} {\bibinfo  {journal} {Phys. Rev. B}\ }\textbf {\bibinfo {volume}
  {105}},\ \bibinfo {pages} {064405} (\bibinfo {year} {2022})}\BibitemShut
  {NoStop}%
\bibitem [{\citenamefont {Bergman}\ \emph {et~al.}(2021)\citenamefont
  {Bergman}, \citenamefont {Duggan}, \citenamefont {Sharma}, \citenamefont
  {Tur}, \citenamefont {Zadok},\ and\ \citenamefont {Al{\`{u}}}}]{Bergman2021}%
  \BibitemOpen
  \bibfield  {author} {\bibinfo {author} {\bibfnamefont {A.}~\bibnamefont
  {Bergman}}, \bibinfo {author} {\bibfnamefont {R.}~\bibnamefont {Duggan}},
  \bibinfo {author} {\bibfnamefont {K.}~\bibnamefont {Sharma}}, \bibinfo
  {author} {\bibfnamefont {M.}~\bibnamefont {Tur}}, \bibinfo {author}
  {\bibfnamefont {A.}~\bibnamefont {Zadok}},\ and\ \bibinfo {author}
  {\bibfnamefont {A.}~\bibnamefont {Al{\`{u}}}},\ }\href
  {https://doi.org/10.1038/s41467-020-20797-7} {\bibfield  {journal} {\bibinfo
  {journal} {Nat. Commun.}\ }\textbf {\bibinfo {volume} {12}},\ \bibinfo
  {pages} {486} (\bibinfo {year} {2021})}\BibitemShut {NoStop}%
\bibitem [{\citenamefont {Antonosyan}\ \emph {et~al.}(2015)\citenamefont
  {Antonosyan}, \citenamefont {Solntsev},\ and\ \citenamefont
  {Sukhorukov}}]{Antonosyan2015}%
  \BibitemOpen
  \bibfield  {author} {\bibinfo {author} {\bibfnamefont {D.~A.}\ \bibnamefont
  {Antonosyan}}, \bibinfo {author} {\bibfnamefont {A.~S.}\ \bibnamefont
  {Solntsev}},\ and\ \bibinfo {author} {\bibfnamefont {A.~A.}\ \bibnamefont
  {Sukhorukov}},\ }\href {https://doi.org/10.1364/OL.40.004575} {\bibfield
  {journal} {\bibinfo  {journal} {Opt. Lett.}\ }\textbf {\bibinfo {volume}
  {40}},\ \bibinfo {pages} {4575} (\bibinfo {year} {2015})}\BibitemShut
  {NoStop}%
\bibitem [{\citenamefont {Li}\ \emph {et~al.}(2021)\citenamefont {Li},
  \citenamefont {Zhou}, \citenamefont {Han}, \citenamefont {Chang},
  \citenamefont {Jiang}, \citenamefont {Huang}, \citenamefont {Zhang},\ and\
  \citenamefont {Xiao}}]{Li2021}%
  \BibitemOpen
  \bibfield  {author} {\bibinfo {author} {\bibfnamefont {W.}~\bibnamefont
  {Li}}, \bibinfo {author} {\bibfnamefont {Y.}~\bibnamefont {Zhou}}, \bibinfo
  {author} {\bibfnamefont {P.}~\bibnamefont {Han}}, \bibinfo {author}
  {\bibfnamefont {X.}~\bibnamefont {Chang}}, \bibinfo {author} {\bibfnamefont
  {S.}~\bibnamefont {Jiang}}, \bibinfo {author} {\bibfnamefont
  {A.}~\bibnamefont {Huang}}, \bibinfo {author} {\bibfnamefont
  {H.}~\bibnamefont {Zhang}},\ and\ \bibinfo {author} {\bibfnamefont
  {Z.}~\bibnamefont {Xiao}},\ }\href
  {https://doi.org/10.1103/PhysRevA.104.033505} {\bibfield  {journal} {\bibinfo
   {journal} {Phys. Rev. A}\ }\textbf {\bibinfo {volume} {104}},\ \bibinfo
  {pages} {033505} (\bibinfo {year} {2021})}\BibitemShut {NoStop}%
\bibitem [{\citenamefont {Blais}\ \emph {et~al.}(2021)\citenamefont {Blais},
  \citenamefont {Grimsmo}, \citenamefont {Girvin},\ and\ \citenamefont
  {Wallraff}}]{Blais2021}%
  \BibitemOpen
  \bibfield  {author} {\bibinfo {author} {\bibfnamefont {A.}~\bibnamefont
  {Blais}}, \bibinfo {author} {\bibfnamefont {A.~L.}\ \bibnamefont {Grimsmo}},
  \bibinfo {author} {\bibfnamefont {S.~M.}\ \bibnamefont {Girvin}},\ and\
  \bibinfo {author} {\bibfnamefont {A.}~\bibnamefont {Wallraff}},\ }\href
  {https://doi.org/10.1103/RevModPhys.93.025005} {\bibfield  {journal}
  {\bibinfo  {journal} {Rev. Mod. Phys.}\ }\textbf {\bibinfo {volume} {93}},\
  \bibinfo {pages} {25005} (\bibinfo {year} {2021})}\BibitemShut {NoStop}%
\bibitem [{\citenamefont {Gothe}\ \emph {et~al.}(2019)\citenamefont {Gothe},
  \citenamefont {Valenzuela}, \citenamefont {Cristiani},\ and\ \citenamefont
  {Eschner}}]{Gothe2019}%
  \BibitemOpen
  \bibfield  {author} {\bibinfo {author} {\bibfnamefont {H.}~\bibnamefont
  {Gothe}}, \bibinfo {author} {\bibfnamefont {T.}~\bibnamefont {Valenzuela}},
  \bibinfo {author} {\bibfnamefont {M.}~\bibnamefont {Cristiani}},\ and\
  \bibinfo {author} {\bibfnamefont {J.}~\bibnamefont {Eschner}},\ }\href
  {https://doi.org/10.1103/PhysRevA.99.013849} {\bibfield  {journal} {\bibinfo
  {journal} {Phys. Rev. A}\ }\textbf {\bibinfo {volume} {99}},\ \bibinfo
  {pages} {013849} (\bibinfo {year} {2019})}\BibitemShut {NoStop}%
\bibitem [{\citenamefont {Fonseca}\ \emph {et~al.}(2016)\citenamefont
  {Fonseca}, \citenamefont {Aranas}, \citenamefont {Millen}, \citenamefont
  {Monteiro},\ and\ \citenamefont {Barker}}]{Fonseca2016}%
  \BibitemOpen
  \bibfield  {author} {\bibinfo {author} {\bibfnamefont {P.~Z.}\ \bibnamefont
  {Fonseca}}, \bibinfo {author} {\bibfnamefont {E.~B.}\ \bibnamefont {Aranas}},
  \bibinfo {author} {\bibfnamefont {J.}~\bibnamefont {Millen}}, \bibinfo
  {author} {\bibfnamefont {T.~S.}\ \bibnamefont {Monteiro}},\ and\ \bibinfo
  {author} {\bibfnamefont {P.~F.}\ \bibnamefont {Barker}},\ }\href
  {https://doi.org/10.1103/PhysRevLett.117.173602} {\bibfield  {journal}
  {\bibinfo  {journal} {Phys. Rev. Lett.}\ }\textbf {\bibinfo {volume} {117}},\
  \bibinfo {pages} {173602} (\bibinfo {year} {2016})}\BibitemShut {NoStop}%
\bibitem [{\citenamefont {Zheng}\ \emph {et~al.}(2020)\citenamefont {Zheng},
  \citenamefont {Zhou}, \citenamefont {Dong}, \citenamefont {Qiu},
  \citenamefont {Chen}, \citenamefont {Guo},\ and\ \citenamefont
  {Sun}}]{Zheng2020}%
  \BibitemOpen
  \bibfield  {author} {\bibinfo {author} {\bibfnamefont {Y.}~\bibnamefont
  {Zheng}}, \bibinfo {author} {\bibfnamefont {L.~M.}\ \bibnamefont {Zhou}},
  \bibinfo {author} {\bibfnamefont {Y.}~\bibnamefont {Dong}}, \bibinfo {author}
  {\bibfnamefont {C.~W.}\ \bibnamefont {Qiu}}, \bibinfo {author} {\bibfnamefont
  {X.~D.}\ \bibnamefont {Chen}}, \bibinfo {author} {\bibfnamefont {G.~C.}\
  \bibnamefont {Guo}},\ and\ \bibinfo {author} {\bibfnamefont {F.~W.}\
  \bibnamefont {Sun}},\ }\href {https://doi.org/10.1103/PhysRevLett.124.223603}
  {\bibfield  {journal} {\bibinfo  {journal} {Phys. Rev. Lett.}\ }\textbf
  {\bibinfo {volume} {124}},\ \bibinfo {pages} {223603} (\bibinfo {year}
  {2020})}\BibitemShut {NoStop}%
\bibitem [{\citenamefont {Katz}\ \emph {et~al.}(2007)\citenamefont {Katz},
  \citenamefont {Retzker}, \citenamefont {Straub},\ and\ \citenamefont
  {Lifshitz}}]{Katz2007}%
  \BibitemOpen
  \bibfield  {author} {\bibinfo {author} {\bibfnamefont {I.}~\bibnamefont
  {Katz}}, \bibinfo {author} {\bibfnamefont {A.}~\bibnamefont {Retzker}},
  \bibinfo {author} {\bibfnamefont {R.}~\bibnamefont {Straub}},\ and\ \bibinfo
  {author} {\bibfnamefont {R.}~\bibnamefont {Lifshitz}},\ }\href
  {https://doi.org/10.1103/PhysRevLett.99.040404} {\bibfield  {journal}
  {\bibinfo  {journal} {Phys. Rev. Lett.}\ }\textbf {\bibinfo {volume} {99}},\
  \bibinfo {pages} {040404} (\bibinfo {year} {2007})}\BibitemShut {NoStop}%
\bibitem [{\citenamefont {Aspelmeyer}\ \emph {et~al.}(2013)\citenamefont
  {Aspelmeyer}, \citenamefont {Kippenberg},\ and\ \citenamefont
  {Marquardt}}]{Aspelmeyer2013}%
  \BibitemOpen
  \bibfield  {author} {\bibinfo {author} {\bibfnamefont {M.}~\bibnamefont
  {Aspelmeyer}}, \bibinfo {author} {\bibfnamefont {T.~J.}\ \bibnamefont
  {Kippenberg}},\ and\ \bibinfo {author} {\bibfnamefont {F.}~\bibnamefont
  {Marquardt}},\ }\href {https://doi.org/10.1103/RevModPhys.86.1391} {\bibfield
   {journal} {\bibinfo  {journal} {Rev. Mod. Phys.}\ }\textbf {\bibinfo
  {volume} {86}},\ \bibinfo {pages} {1391} (\bibinfo {year}
  {2013})}\BibitemShut {NoStop}%
\bibitem [{\citenamefont {Nair}\ \emph
  {et~al.}(2021{\natexlab{b}})\citenamefont {Nair}, \citenamefont
  {Mukhopadhyay},\ and\ \citenamefont {Agarwal}}]{Nair2021b}%
  \BibitemOpen
  \bibfield  {author} {\bibinfo {author} {\bibfnamefont {J.~M.~P.}\
  \bibnamefont {Nair}}, \bibinfo {author} {\bibfnamefont {D.}~\bibnamefont
  {Mukhopadhyay}},\ and\ \bibinfo {author} {\bibfnamefont {G.~S.}\ \bibnamefont
  {Agarwal}},\ }\href {https://doi.org/10.1103/PhysRevB.103.224401} {\bibfield
  {journal} {\bibinfo  {journal} {Phys. Rev. B}\ }\textbf {\bibinfo {volume}
  {103}},\ \bibinfo {pages} {224401} (\bibinfo {year}
  {2021}{\natexlab{b}})}\BibitemShut {NoStop}%
\bibitem [{\citenamefont {Juan}\ \emph {et~al.}(2016)\citenamefont {Juan},
  \citenamefont {Molina-Terriza}, \citenamefont {Volz},\ and\ \citenamefont
  {Romero-Isart}}]{Juan2016}%
  \BibitemOpen
  \bibfield  {author} {\bibinfo {author} {\bibfnamefont {M.~L.}\ \bibnamefont
  {Juan}}, \bibinfo {author} {\bibfnamefont {G.}~\bibnamefont
  {Molina-Terriza}}, \bibinfo {author} {\bibfnamefont {T.}~\bibnamefont
  {Volz}},\ and\ \bibinfo {author} {\bibfnamefont {O.}~\bibnamefont
  {Romero-Isart}},\ }\href {https://doi.org/10.1103/PhysRevA.94.023841}
  {\bibfield  {journal} {\bibinfo  {journal} {Phys. Rev. A}\ }\textbf {\bibinfo
  {volume} {94}},\ \bibinfo {pages} {023841} (\bibinfo {year}
  {2016})}\BibitemShut {NoStop}%
\bibitem [{\citenamefont {Harder}\ \emph {et~al.}(2018)\citenamefont {Harder},
  \citenamefont {Yang}, \citenamefont {Yao}, \citenamefont {Yu}, \citenamefont
  {Rao}, \citenamefont {Gui}, \citenamefont {Stamps},\ and\ \citenamefont
  {Hu}}]{Harder2018}%
  \BibitemOpen
  \bibfield  {author} {\bibinfo {author} {\bibfnamefont {M.}~\bibnamefont
  {Harder}}, \bibinfo {author} {\bibfnamefont {Y.}~\bibnamefont {Yang}},
  \bibinfo {author} {\bibfnamefont {B.~M.}\ \bibnamefont {Yao}}, \bibinfo
  {author} {\bibfnamefont {C.~H.}\ \bibnamefont {Yu}}, \bibinfo {author}
  {\bibfnamefont {J.~W.}\ \bibnamefont {Rao}}, \bibinfo {author} {\bibfnamefont
  {Y.~S.}\ \bibnamefont {Gui}}, \bibinfo {author} {\bibfnamefont {R.~L.}\
  \bibnamefont {Stamps}},\ and\ \bibinfo {author} {\bibfnamefont {C.~M.}\
  \bibnamefont {Hu}},\ }\href {https://doi.org/10.1103/PhysRevLett.121.137203}
  {\bibfield  {journal} {\bibinfo  {journal} {Phys. Rev. Lett.}\ }\textbf
  {\bibinfo {volume} {121}},\ \bibinfo {pages} {137203} (\bibinfo {year}
  {2018})}\BibitemShut {NoStop}%
\bibitem [{\citenamefont {{Zare Rameshti}}\ and\ \citenamefont
  {Bauer}(2018)}]{ZareRameshti2018}%
  \BibitemOpen
  \bibfield  {author} {\bibinfo {author} {\bibfnamefont {B.}~\bibnamefont
  {{Zare Rameshti}}}\ and\ \bibinfo {author} {\bibfnamefont {G.~E.}\
  \bibnamefont {Bauer}},\ }\href {https://doi.org/10.1103/PhysRevB.97.014419}
  {\bibfield  {journal} {\bibinfo  {journal} {Phys. Rev. B}\ }\textbf {\bibinfo
  {volume} {97}},\ \bibinfo {pages} {014419} (\bibinfo {year}
  {2018})}\BibitemShut {NoStop}%
\bibitem [{\citenamefont {Fang}\ \emph {et~al.}(2016)\citenamefont {Fang},
  \citenamefont {Matheny}, \citenamefont {Luan},\ and\ \citenamefont
  {Painter}}]{Fang2016}%
  \BibitemOpen
  \bibfield  {author} {\bibinfo {author} {\bibfnamefont {K.}~\bibnamefont
  {Fang}}, \bibinfo {author} {\bibfnamefont {M.~H.}\ \bibnamefont {Matheny}},
  \bibinfo {author} {\bibfnamefont {X.}~\bibnamefont {Luan}},\ and\ \bibinfo
  {author} {\bibfnamefont {O.}~\bibnamefont {Painter}},\ }\href
  {https://doi.org/10.1038/nphoton.2016.107} {\bibfield  {journal} {\bibinfo
  {journal} {Nat. Photonics}\ }\textbf {\bibinfo {volume} {10}},\ \bibinfo
  {pages} {489} (\bibinfo {year} {2016})}\BibitemShut {NoStop}%
\bibitem [{\citenamefont {Bernier}\ \emph {et~al.}(2018)\citenamefont
  {Bernier}, \citenamefont {T{\'{o}}th}, \citenamefont {Feofanov},\ and\
  \citenamefont {Kippenberg}}]{Bernier2018}%
  \BibitemOpen
  \bibfield  {author} {\bibinfo {author} {\bibfnamefont {N.~R.}\ \bibnamefont
  {Bernier}}, \bibinfo {author} {\bibfnamefont {L.~D.}\ \bibnamefont
  {T{\'{o}}th}}, \bibinfo {author} {\bibfnamefont {A.~K.}\ \bibnamefont
  {Feofanov}},\ and\ \bibinfo {author} {\bibfnamefont {T.~J.}\ \bibnamefont
  {Kippenberg}},\ }\href {https://doi.org/10.1103/PhysRevA.98.023841}
  {\bibfield  {journal} {\bibinfo  {journal} {Phys. Rev. A}\ }\textbf {\bibinfo
  {volume} {98}},\ \bibinfo {pages} {023841} (\bibinfo {year}
  {2018})}\BibitemShut {NoStop}%
\bibitem [{\citenamefont {Metelmann}\ and\ \citenamefont
  {Clerk}(2015)}]{Metelmann2015}%
  \BibitemOpen
  \bibfield  {author} {\bibinfo {author} {\bibfnamefont {A.}~\bibnamefont
  {Metelmann}}\ and\ \bibinfo {author} {\bibfnamefont {A.~A.}\ \bibnamefont
  {Clerk}},\ }\href {https://doi.org/10.1103/PhysRevX.5.021025} {\bibfield
  {journal} {\bibinfo  {journal} {Phys. Rev. X}\ }\textbf {\bibinfo {volume}
  {5}},\ \bibinfo {pages} {021025} (\bibinfo {year} {2015})}\BibitemShut
  {NoStop}%
\bibitem [{\citenamefont {Chang}\ \emph {et~al.}(2011)\citenamefont {Chang},
  \citenamefont {Safavi-Naeini}, \citenamefont {Hafezi},\ and\ \citenamefont
  {Painter}}]{Chang2011}%
  \BibitemOpen
  \bibfield  {author} {\bibinfo {author} {\bibfnamefont {D.~E.}\ \bibnamefont
  {Chang}}, \bibinfo {author} {\bibfnamefont {A.~H.}\ \bibnamefont
  {Safavi-Naeini}}, \bibinfo {author} {\bibfnamefont {M.}~\bibnamefont
  {Hafezi}},\ and\ \bibinfo {author} {\bibfnamefont {O.}~\bibnamefont
  {Painter}},\ }\href {https://doi.org/10.1088/1367-2630/13/2/023003}
  {\bibfield  {journal} {\bibinfo  {journal} {New J. Phys.}\ }\textbf {\bibinfo
  {volume} {13}},\ \bibinfo {pages} {023003} (\bibinfo {year}
  {2011})}\BibitemShut {NoStop}%
\bibitem [{\citenamefont {Wang}\ \emph {et~al.}(2019)\citenamefont {Wang},
  \citenamefont {Du}, \citenamefont {Li},\ and\ \citenamefont
  {Liu}}]{Wang2019}%
  \BibitemOpen
  \bibfield  {author} {\bibinfo {author} {\bibfnamefont {Z.}~\bibnamefont
  {Wang}}, \bibinfo {author} {\bibfnamefont {L.}~\bibnamefont {Du}}, \bibinfo
  {author} {\bibfnamefont {Y.}~\bibnamefont {Li}},\ and\ \bibinfo {author}
  {\bibfnamefont {Y.~X.}\ \bibnamefont {Liu}},\ }\href
  {https://doi.org/10.1103/PhysRevA.100.053809} {\bibfield  {journal} {\bibinfo
   {journal} {Phys. Rev. A}\ }\textbf {\bibinfo {volume} {100}},\ \bibinfo
  {pages} {053809} (\bibinfo {year} {2019})}\BibitemShut {NoStop}%
\bibitem [{\citenamefont {Du}\ \emph {et~al.}(2021{\natexlab{a}})\citenamefont
  {Du}, \citenamefont {Chen},\ and\ \citenamefont {Li}}]{Du2021}%
  \BibitemOpen
  \bibfield  {author} {\bibinfo {author} {\bibfnamefont {L.}~\bibnamefont
  {Du}}, \bibinfo {author} {\bibfnamefont {Y.~T.}\ \bibnamefont {Chen}},\ and\
  \bibinfo {author} {\bibfnamefont {Y.}~\bibnamefont {Li}},\ }\href
  {https://doi.org/10.1103/PhysRevResearch.3.043226} {\bibfield  {journal}
  {\bibinfo  {journal} {Phys. Rev. Research}\ }\textbf {\bibinfo {volume}
  {3}},\ \bibinfo {pages} {043226} (\bibinfo {year}
  {2021}{\natexlab{a}})}\BibitemShut {NoStop}%
\bibitem [{\citenamefont {Xiao}\ \emph {et~al.}(2010)\citenamefont {Xiao},
  \citenamefont {Li}, \citenamefont {Liu}, \citenamefont {Li}, \citenamefont
  {Sun},\ and\ \citenamefont {Gong}}]{Xiao2010}%
  \BibitemOpen
  \bibfield  {author} {\bibinfo {author} {\bibfnamefont {Y.~F.}\ \bibnamefont
  {Xiao}}, \bibinfo {author} {\bibfnamefont {M.}~\bibnamefont {Li}}, \bibinfo
  {author} {\bibfnamefont {Y.~C.}\ \bibnamefont {Liu}}, \bibinfo {author}
  {\bibfnamefont {Y.}~\bibnamefont {Li}}, \bibinfo {author} {\bibfnamefont
  {X.}~\bibnamefont {Sun}},\ and\ \bibinfo {author} {\bibfnamefont
  {Q.}~\bibnamefont {Gong}},\ }\href
  {https://doi.org/10.1103/PhysRevA.82.065804} {\bibfield  {journal} {\bibinfo
  {journal} {Phys. Rev. A}\ }\textbf {\bibinfo {volume} {82}},\ \bibinfo
  {pages} {065804} (\bibinfo {year} {2010})}\BibitemShut {NoStop}%
\bibitem [{\citenamefont {Du}\ \emph {et~al.}(2021{\natexlab{b}})\citenamefont
  {Du}, \citenamefont {Wang},\ and\ \citenamefont {Li}}]{Du2021oe}%
  \BibitemOpen
  \bibfield  {author} {\bibinfo {author} {\bibfnamefont {L.}~\bibnamefont
  {Du}}, \bibinfo {author} {\bibfnamefont {Z.}~\bibnamefont {Wang}},\ and\
  \bibinfo {author} {\bibfnamefont {Y.}~\bibnamefont {Li}},\ }\href
  {https://doi.org/10.1364/oe.412996} {\bibfield  {journal} {\bibinfo
  {journal} {Opt. Express}\ }\textbf {\bibinfo {volume} {29}},\ \bibinfo
  {pages} {3038} (\bibinfo {year} {2021}{\natexlab{b}})}\BibitemShut {NoStop}%
\bibitem [{\citenamefont {Arcizet}\ \emph {et~al.}(2006)\citenamefont
  {Arcizet}, \citenamefont {Cohadon}, \citenamefont {Briant}, \citenamefont
  {Pinard}, \citenamefont {Heidmann}, \citenamefont {MacKowski}, \citenamefont
  {Michel}, \citenamefont {Pinard}, \citenamefont {Fran{\c{c}}ais},\ and\
  \citenamefont {Rousseau}}]{Arcizet2006}%
  \BibitemOpen
  \bibfield  {author} {\bibinfo {author} {\bibfnamefont {O.}~\bibnamefont
  {Arcizet}}, \bibinfo {author} {\bibfnamefont {P.~F.}\ \bibnamefont
  {Cohadon}}, \bibinfo {author} {\bibfnamefont {T.}~\bibnamefont {Briant}},
  \bibinfo {author} {\bibfnamefont {M.}~\bibnamefont {Pinard}}, \bibinfo
  {author} {\bibfnamefont {A.}~\bibnamefont {Heidmann}}, \bibinfo {author}
  {\bibfnamefont {J.~M.}\ \bibnamefont {MacKowski}}, \bibinfo {author}
  {\bibfnamefont {C.}~\bibnamefont {Michel}}, \bibinfo {author} {\bibfnamefont
  {L.}~\bibnamefont {Pinard}}, \bibinfo {author} {\bibfnamefont
  {O.}~\bibnamefont {Fran{\c{c}}ais}},\ and\ \bibinfo {author} {\bibfnamefont
  {L.}~\bibnamefont {Rousseau}},\ }\href
  {https://doi.org/10.1103/PhysRevLett.97.133601} {\bibfield  {journal}
  {\bibinfo  {journal} {Phys. Rev. Lett.}\ }\textbf {\bibinfo {volume} {97}},\
  \bibinfo {pages} {133601} (\bibinfo {year} {2006})}\BibitemShut {NoStop}%
\bibitem [{\citenamefont {Gr{\"{o}}blacher}\ \emph
  {et~al.}(2009{\natexlab{a}})\citenamefont {Gr{\"{o}}blacher}, \citenamefont
  {Hertzberg}, \citenamefont {Vanner}, \citenamefont {Cole}, \citenamefont
  {Gigan}, \citenamefont {Schwab},\ and\ \citenamefont
  {Aspelmeyer}}]{Groeblacher2009}%
  \BibitemOpen
  \bibfield  {author} {\bibinfo {author} {\bibfnamefont {S.}~\bibnamefont
  {Gr{\"{o}}blacher}}, \bibinfo {author} {\bibfnamefont {J.~B.}\ \bibnamefont
  {Hertzberg}}, \bibinfo {author} {\bibfnamefont {M.~R.}\ \bibnamefont
  {Vanner}}, \bibinfo {author} {\bibfnamefont {G.~D.}\ \bibnamefont {Cole}},
  \bibinfo {author} {\bibfnamefont {S.}~\bibnamefont {Gigan}}, \bibinfo
  {author} {\bibfnamefont {K.~C.}\ \bibnamefont {Schwab}},\ and\ \bibinfo
  {author} {\bibfnamefont {M.}~\bibnamefont {Aspelmeyer}},\ }\href
  {https://doi.org/10.1038/nphys1301} {\bibfield  {journal} {\bibinfo
  {journal} {Nat. Phys.}\ }\textbf {\bibinfo {volume} {5}},\ \bibinfo {pages}
  {485} (\bibinfo {year} {2009}{\natexlab{a}})}\BibitemShut {NoStop}%
\bibitem [{\citenamefont {Gr{\"{o}}blacher}\ \emph
  {et~al.}(2009{\natexlab{b}})\citenamefont {Gr{\"{o}}blacher}, \citenamefont
  {Hammerer}, \citenamefont {Vanner},\ and\ \citenamefont
  {Aspelmeyer}}]{Groeblacher2009a}%
  \BibitemOpen
  \bibfield  {author} {\bibinfo {author} {\bibfnamefont {S.}~\bibnamefont
  {Gr{\"{o}}blacher}}, \bibinfo {author} {\bibfnamefont {K.}~\bibnamefont
  {Hammerer}}, \bibinfo {author} {\bibfnamefont {M.~R.}\ \bibnamefont
  {Vanner}},\ and\ \bibinfo {author} {\bibfnamefont {M.}~\bibnamefont
  {Aspelmeyer}},\ }\href {https://doi.org/10.1038/nature08171} {\bibfield
  {journal} {\bibinfo  {journal} {Nature (London)}\ }\textbf {\bibinfo {volume}
  {460}},\ \bibinfo {pages} {724} (\bibinfo {year}
  {2009}{\natexlab{b}})}\BibitemShut {NoStop}%
\bibitem [{\citenamefont {Kleckner}\ \emph {et~al.}(2011)\citenamefont
  {Kleckner}, \citenamefont {Pepper}, \citenamefont {Jeffrey}, \citenamefont
  {Sonin}, \citenamefont {Thon},\ and\ \citenamefont
  {Bouwmeester}}]{Kleckner2011}%
  \BibitemOpen
  \bibfield  {author} {\bibinfo {author} {\bibfnamefont {D.}~\bibnamefont
  {Kleckner}}, \bibinfo {author} {\bibfnamefont {B.}~\bibnamefont {Pepper}},
  \bibinfo {author} {\bibfnamefont {E.}~\bibnamefont {Jeffrey}}, \bibinfo
  {author} {\bibfnamefont {P.}~\bibnamefont {Sonin}}, \bibinfo {author}
  {\bibfnamefont {S.~M.}\ \bibnamefont {Thon}},\ and\ \bibinfo {author}
  {\bibfnamefont {D.}~\bibnamefont {Bouwmeester}},\ }\href
  {https://doi.org/10.1364/oe.19.019708} {\bibfield  {journal} {\bibinfo
  {journal} {Opt. Express}\ }\textbf {\bibinfo {volume} {19}},\ \bibinfo
  {pages} {19708} (\bibinfo {year} {2011})}\BibitemShut {NoStop}%
\bibitem [{\citenamefont {Rocheleau}\ \emph {et~al.}(2010)\citenamefont
  {Rocheleau}, \citenamefont {Ndukum}, \citenamefont {MacKlin}, \citenamefont
  {Hertzberg}, \citenamefont {Clerk},\ and\ \citenamefont
  {Schwab}}]{Rocheleau2010}%
  \BibitemOpen
  \bibfield  {author} {\bibinfo {author} {\bibfnamefont {T.}~\bibnamefont
  {Rocheleau}}, \bibinfo {author} {\bibfnamefont {T.}~\bibnamefont {Ndukum}},
  \bibinfo {author} {\bibfnamefont {C.}~\bibnamefont {MacKlin}}, \bibinfo
  {author} {\bibfnamefont {J.~B.}\ \bibnamefont {Hertzberg}}, \bibinfo {author}
  {\bibfnamefont {A.~A.}\ \bibnamefont {Clerk}},\ and\ \bibinfo {author}
  {\bibfnamefont {K.~C.}\ \bibnamefont {Schwab}},\ }\href
  {https://doi.org/10.1038/nature08681} {\bibfield  {journal} {\bibinfo
  {journal} {Nature (London)}\ }\textbf {\bibinfo {volume} {463}},\ \bibinfo
  {pages} {72} (\bibinfo {year} {2010})}\BibitemShut {NoStop}%
\bibitem [{\citenamefont {Shi}\ and\ \citenamefont {Fan}(2013)}]{Shi2013}%
  \BibitemOpen
  \bibfield  {author} {\bibinfo {author} {\bibfnamefont {T.}~\bibnamefont
  {Shi}}\ and\ \bibinfo {author} {\bibfnamefont {S.}~\bibnamefont {Fan}},\
  }\href {https://doi.org/10.1103/PhysRevA.87.063818} {\bibfield  {journal}
  {\bibinfo  {journal} {Phys. Rev. A}\ }\textbf {\bibinfo {volume} {87}},\
  \bibinfo {pages} {063818} (\bibinfo {year} {2013})}\BibitemShut {NoStop}%
\bibitem [{\citenamefont {Dayan}\ \emph {et~al.}(2008)\citenamefont {Dayan},
  \citenamefont {Parkins}, \citenamefont {Aoki}, \citenamefont {Ostby},
  \citenamefont {Vahala},\ and\ \citenamefont {Kimble}}]{Dayan2008}%
  \BibitemOpen
  \bibfield  {author} {\bibinfo {author} {\bibfnamefont {B.}~\bibnamefont
  {Dayan}}, \bibinfo {author} {\bibfnamefont {a.~S.}\ \bibnamefont {Parkins}},
  \bibinfo {author} {\bibfnamefont {T.}~\bibnamefont {Aoki}}, \bibinfo {author}
  {\bibfnamefont {E.~P.}\ \bibnamefont {Ostby}}, \bibinfo {author}
  {\bibfnamefont {K.~J.}\ \bibnamefont {Vahala}},\ and\ \bibinfo {author}
  {\bibfnamefont {H.~J.}\ \bibnamefont {Kimble}},\ }\href@noop {} {\bibfield
  {journal} {\bibinfo  {journal} {Science}\ }\textbf {\bibinfo {volume}
  {319}},\ \bibinfo {pages} {22} (\bibinfo {year} {2008})}\BibitemShut
  {NoStop}%
\bibitem [{\citenamefont {Takashima}\ \emph {et~al.}(2014)\citenamefont
  {Takashima}, \citenamefont {Kitajima}, \citenamefont {Tanaka}, \citenamefont
  {Fujiwara},\ and\ \citenamefont {Sasaki}}]{Takashima2014}%
  \BibitemOpen
  \bibfield  {author} {\bibinfo {author} {\bibfnamefont {H.}~\bibnamefont
  {Takashima}}, \bibinfo {author} {\bibfnamefont {K.}~\bibnamefont {Kitajima}},
  \bibinfo {author} {\bibfnamefont {Y.}~\bibnamefont {Tanaka}}, \bibinfo
  {author} {\bibfnamefont {H.}~\bibnamefont {Fujiwara}},\ and\ \bibinfo
  {author} {\bibfnamefont {K.}~\bibnamefont {Sasaki}},\ }\href
  {https://doi.org/10.1103/PhysRevA.89.021801} {\bibfield  {journal} {\bibinfo
  {journal} {Phys. Rev. A}\ }\textbf {\bibinfo {volume} {89}},\ \bibinfo
  {pages} {021801} (\bibinfo {year} {2014})}\BibitemShut {NoStop}%
\bibitem [{\citenamefont {Neukirch}\ \emph {et~al.}(2013)\citenamefont
  {Neukirch}, \citenamefont {Gieseler}, \citenamefont {Quidant}, \citenamefont
  {Novotny},\ and\ \citenamefont {{Nick Vamivakas}}}]{Neukirch2013}%
  \BibitemOpen
  \bibfield  {author} {\bibinfo {author} {\bibfnamefont {L.~P.}\ \bibnamefont
  {Neukirch}}, \bibinfo {author} {\bibfnamefont {J.}~\bibnamefont {Gieseler}},
  \bibinfo {author} {\bibfnamefont {R.}~\bibnamefont {Quidant}}, \bibinfo
  {author} {\bibfnamefont {L.}~\bibnamefont {Novotny}},\ and\ \bibinfo {author}
  {\bibfnamefont {A.}~\bibnamefont {{Nick Vamivakas}}},\ }\href
  {https://doi.org/10.1364/OL.38.002976} {\bibfield  {journal} {\bibinfo
  {journal} {Opt. Lett.}\ }\textbf {\bibinfo {volume} {38}},\ \bibinfo {pages}
  {2976} (\bibinfo {year} {2013})}\BibitemShut {NoStop}%
\bibitem [{\citenamefont {Neukirch}\ \emph {et~al.}(2015)\citenamefont
  {Neukirch}, \citenamefont {von Haartman}, \citenamefont {Rosenholm},\ and\
  \citenamefont {{Nick Vamivakas}}}]{Neukirch2015}%
  \BibitemOpen
  \bibfield  {author} {\bibinfo {author} {\bibfnamefont {L.~P.}\ \bibnamefont
  {Neukirch}}, \bibinfo {author} {\bibfnamefont {E.}~\bibnamefont {von
  Haartman}}, \bibinfo {author} {\bibfnamefont {J.~M.}\ \bibnamefont
  {Rosenholm}},\ and\ \bibinfo {author} {\bibfnamefont {A.}~\bibnamefont {{Nick
  Vamivakas}}},\ }\href {https://doi.org/10.1038/nphoton.2015.162} {\bibfield
  {journal} {\bibinfo  {journal} {Nat. Photonics}\ }\textbf {\bibinfo {volume}
  {9}},\ \bibinfo {pages} {653} (\bibinfo {year} {2015})}\BibitemShut {NoStop}%
\bibitem [{\citenamefont {Hoang}\ \emph {et~al.}(2016)\citenamefont {Hoang},
  \citenamefont {Ahn}, \citenamefont {Bang},\ and\ \citenamefont
  {Li}}]{Hoang2016}%
  \BibitemOpen
  \bibfield  {author} {\bibinfo {author} {\bibfnamefont {T.~M.}\ \bibnamefont
  {Hoang}}, \bibinfo {author} {\bibfnamefont {J.}~\bibnamefont {Ahn}}, \bibinfo
  {author} {\bibfnamefont {J.}~\bibnamefont {Bang}},\ and\ \bibinfo {author}
  {\bibfnamefont {T.}~\bibnamefont {Li}},\ }\href
  {https://doi.org/10.1038/ncomms12250} {\bibfield  {journal} {\bibinfo
  {journal} {Nat. Commun.}\ }\textbf {\bibinfo {volume} {7}},\ \bibinfo {pages}
  {12250} (\bibinfo {year} {2016})}\BibitemShut {NoStop}%
\bibitem [{\citenamefont {Chang}\ \emph {et~al.}(2010)\citenamefont {Chang},
  \citenamefont {Regal}, \citenamefont {Papp}, \citenamefont {Wilson},
  \citenamefont {Ye}, \citenamefont {Painter}, \citenamefont {Kimble},\ and\
  \citenamefont {Zoller}}]{Chang2010}%
  \BibitemOpen
  \bibfield  {author} {\bibinfo {author} {\bibfnamefont {D.~E.}\ \bibnamefont
  {Chang}}, \bibinfo {author} {\bibfnamefont {C.~A.}\ \bibnamefont {Regal}},
  \bibinfo {author} {\bibfnamefont {S.~B.}\ \bibnamefont {Papp}}, \bibinfo
  {author} {\bibfnamefont {D.~J.}\ \bibnamefont {Wilson}}, \bibinfo {author}
  {\bibfnamefont {J.}~\bibnamefont {Ye}}, \bibinfo {author} {\bibfnamefont
  {O.}~\bibnamefont {Painter}}, \bibinfo {author} {\bibfnamefont {H.~J.}\
  \bibnamefont {Kimble}},\ and\ \bibinfo {author} {\bibfnamefont
  {P.}~\bibnamefont {Zoller}},\ }\href
  {https://doi.org/10.1073/pnas.0912969107} {\bibfield  {journal} {\bibinfo
  {journal} {Proc. Natl. Acad. Sci. USA}\ }\textbf {\bibinfo {volume} {107}},\
  \bibinfo {pages} {1005} (\bibinfo {year} {2010})}\BibitemShut {NoStop}%
\bibitem [{\citenamefont {Farnesi}\ \emph {et~al.}(2021)\citenamefont
  {Farnesi}, \citenamefont {Pelli}, \citenamefont {Soria}, \citenamefont
  {{Nunzi Conti}}, \citenamefont {{Le Roux}}, \citenamefont {{Montesinos
  Ballester}}, \citenamefont {Vivien}, \citenamefont {Cheben},\ and\
  \citenamefont {Alonso-Ramos}}]{Farnesi2021}%
  \BibitemOpen
  \bibfield  {author} {\bibinfo {author} {\bibfnamefont {D.}~\bibnamefont
  {Farnesi}}, \bibinfo {author} {\bibfnamefont {S.}~\bibnamefont {Pelli}},
  \bibinfo {author} {\bibfnamefont {S.}~\bibnamefont {Soria}}, \bibinfo
  {author} {\bibfnamefont {G.}~\bibnamefont {{Nunzi Conti}}}, \bibinfo {author}
  {\bibfnamefont {X.}~\bibnamefont {{Le Roux}}}, \bibinfo {author}
  {\bibfnamefont {M.}~\bibnamefont {{Montesinos Ballester}}}, \bibinfo {author}
  {\bibfnamefont {L.}~\bibnamefont {Vivien}}, \bibinfo {author} {\bibfnamefont
  {P.}~\bibnamefont {Cheben}},\ and\ \bibinfo {author} {\bibfnamefont
  {C.}~\bibnamefont {Alonso-Ramos}},\ }\href
  {https://doi.org/10.1364/optica.438395} {\bibfield  {journal} {\bibinfo
  {journal} {Optica}\ }\textbf {\bibinfo {volume} {8}},\ \bibinfo {pages}
  {1511} (\bibinfo {year} {2021})}\BibitemShut {NoStop}%
\bibitem [{\citenamefont {Spillane}\ \emph {et~al.}(2003)\citenamefont
  {Spillane}, \citenamefont {Kippenberg}, \citenamefont {Painter},\ and\
  \citenamefont {Vahala}}]{Spillane2003}%
  \BibitemOpen
  \bibfield  {author} {\bibinfo {author} {\bibfnamefont {S.~M.}\ \bibnamefont
  {Spillane}}, \bibinfo {author} {\bibfnamefont {T.~J.}\ \bibnamefont
  {Kippenberg}}, \bibinfo {author} {\bibfnamefont {O.~J.}\ \bibnamefont
  {Painter}},\ and\ \bibinfo {author} {\bibfnamefont {K.~J.}\ \bibnamefont
  {Vahala}},\ }\href {https://doi.org/10.1103/PhysRevLett.91.043902} {\bibfield
   {journal} {\bibinfo  {journal} {Phys. Rev. Lett.}\ }\textbf {\bibinfo
  {volume} {91}},\ \bibinfo {pages} {043902} (\bibinfo {year}
  {2003})}\BibitemShut {NoStop}%
\end{thebibliography}%

\end{document}